\documentclass[showpacs,preprint,amsmath,amssymb,nofootinbib]{revtex4-1}

\usepackage{amsmath}
\usepackage{graphicx}
\usepackage{epsf}
\usepackage{psfrag}
\usepackage{epsfig}
\usepackage{graphics}
\usepackage{amsfonts}
\usepackage{epstopdf}
\usepackage{slashed}
\usepackage{subfigure}

\newcommand{\be}{\begin{equation}}
\newcommand{\ee}{\end{equation}}
\newcommand{\bq}{\begin{eqnarray}}
\newcommand{\eq}{\end{eqnarray}}
\newcommand{\ba}{\begin{align}}
\newcommand{\ea}{\end{align}}

\begin{document}

\title{$\gamma_{5}$ algebra ambiguities in Feynman amplitudes: momentum routing invariance and anomalies in $D=4$ and $D=2$}

\date{\today}

\author{A. C. D. Viglioni$^{(a)}$} \email[]{acdv@fisica.ufmg.br}
\author{A. L. Cherchiglia$^{(b)}$} \email[]{adriano.cherchiglia@mailbox.tu-dresden.de}
\author{A. R. Vieira$^{(a,c)}$} \email[]{arvieira@fisica.ufmg.br}
\author{Brigitte Hiller$^{(d)}$} \email[]{brigitte@teor.fis.uc.pt}
\author{Marcos Sampaio$^{(a)}$} \email []{msampaio@fisica.ufmg.br}

\affiliation{(a) Departamento de F\'isica - ICEx - Universidade Federal de Minas Gerais\\ P.O. BOX 702, 30.161-970, Belo Horizonte - MG - Brasil}
\affiliation{(b) Institut fur Kern- und Teilchenphysik, TU Dresden, Zellescher Weg 19, 01069 Dresden, Germany}
\affiliation{(c) Indiana University Center for Spacetime Symmetries, Bloomington, Indiana 47405, USA}
\affiliation{(d) CFisUC, Department of Physics, University of Coimbra, 3004-516 Coimbra, Portugal}

\begin{abstract}
We address the subject of chiral anomalies in two and four dimensional theories. Ambiguities associated with the $\gamma_5$ algebra within divergent integrals are identified, even though the physical dimension is not altered in the process of regularization. We present a minimal prescription that leads to unique results and apply it to a series of examples. For the particular case of abelian theories with effective chiral vertices, we show: 1- Its implication on the way to display the anomalies democratically in the Ward identities. 2- The possibility to fix an arbitrary surface term in such a way that a momentum routing independent result emerges. This leads to a reinterpretation of the role of momentum routing in the process of choosing the Ward identity to be satisfied in an anomalous process. 3- Momentum Routing Invariance (MRI) is a necessary and sufficient condition to assure vectorial gauge invariance of effective chiral Abelian gauge theories. We also briefly discuss the case of complete chiral theories, using the Chiral Schwinger Model as an example. 
\end{abstract}


\maketitle

\section{Introduction}

The regularization and renormalization program of Quantum Field Theory (QFT) must comply with the regularization scheme independence of theories with anomalously broken symmetries. Anomalies manifest themselves by the impossibility to maintain in the quantum extension of a theory all its classical symmetries intact. As the Ward identity associated with gauge invariance is often required on physical grounds to be the one which is left unbroken, dimensional regularization \cite{{DR},{Bollini}} seemed to be an appropriate method for this case, since it preserves gauge invariance. However, some inconsistencies soon appeared regarding the manipulation of dimension specific objects such as $\gamma_5$ and the Levi-Civita tensor \cite{{Siegel},{Siegel2}}. To circumvent this problem, some rules had to be added to the method, postulating how the dimensional continuation of such objects should be performed \cite{Jegerlehner}. 

The situation is aggravated by the fact, as we are are going to present in this contribution, that even staying in the physical dimension specific to the theory, identities regarding the $\gamma_5$ algebra are not always satisfied under divergent integrals. In other words, the framework to deal with such identities is regularization dependent and, as such, also requires the adoption of some prescription.

Recently, the interest on the subject has been renewed in the literature with new proposals for a gauge-invariant prescription for the $\gamma_5$ algebra. In particular, the author of  \cite{Tsai1,Tsai2} presents the so called Rightmost Ordering in which all $\gamma_5$ should be moved to the rightmost position of the amplitude before its dimensionality is altered. Another proposal focuses on an integral representation for the trace involving gamma matrices \cite{Ferrari1, Ferrari2, Ferrari3}. Nevertheless, in both cases the authors intend to obtain a prescription which allows dimensional regularization to be applied to dimension specific objects as the $\gamma_5$ matrix. Another proposal, in the case of four-dimensional regularization, was envisaged by the authors of \cite{Cynolter}. 

In this contribution we would like to show how anomalies should be consistently dealt with in the Implicit Regularization (IReg) formalism  \cite{IR,BaetaScarpelli:2000zs,Cherchiglia:2010yd}. In IReg no dimensional continuation on space-time is performed and this method possesses the useful property of displaying democratically the Ward identities to be conserved or broken in an anomalous process. This property results from the technique's general treatment of differences of divergent quantities that have the same degree of divergence and are prone to occur in the diagrammatic evaluation of an amplitude. These differences correspond to finite but otherwise arbitrary surface terms. Democracy becomes manifest by keeping these surface terms as arbitrary parameters until the very end of the calculation of an anomalous amplitude. Only then should they be fixed according to the symmetry constraints of the particular physical process one is dealing with. 

Now due to the complications inherent to the $\gamma_5$ algebra inside divergent quantities, ambiguities that may stem from it will be carried over to the arbitrary parameters. With these being arbitrary, it may seem impossible to tell whether the $\gamma_5$ algebra has been performed adequately or not. One of the purposes of the present study is to show that this is not so if one takes advantage of the properties that amplitudes are expected to fulfill under a change in the routing of momenta in the propagators of a loop integral. Momentum routing invariance (MRI) is known to be fulfilled in the cases that a symmetry is not broken at all orders of a theory, such as the gauge symmetries of the Standard Model, upon use of Dimensional Regularization \cite{DR}. It is legitimate to expect that in presence of an anomaly, gauge invariance continues to evidence MRI. Indeed that is what we verify by applying a minimal prescription  based on the symmetrization of the trace over the $\gamma$ matrices involving $\gamma_5$. This prescription does not make use of the property $\{\gamma_5,\gamma_\mu \}=0$, since we show in several examples that the vanishing of the anti-commutator is at the origin of the ambiguities. 

Moreover, adopting this symmetrization prescription, leads to the unforeseen conclusion that independently of the sector which remains unbroken, the vectorial or the chiral, MRI is always verified, i.e. one does not need to fix a particular value of a routing momentum in the process of choosing which Ward identity is to be satisfied. We will show explicitly  for the  Adler Bell Jackiw anomaly that at no instance MRI is broken in the evaluation of the Ward identities. Instead we note that the routing momenta come always accompanied by arbitrary surface terms. It is these that need finally be fixed and this operation can be effected without touching MRI. In other words MRI is protected by the occurrence of the arbitrary surface parameters. The connection between MRI and these arbitrary surface terms has been observed in different contexts \cite{Cherchiglia,Cherchiglia:2012zp,Felipe:2014gma}, the most recent one regarding the study of gauge invariance in extended QED in which case the use of the  symmetrization prescription presented in this contribution was essential to obtain an unambiguous $\gamma_{5}$ algebra \cite{Vieira:2015fra}. 
 
This work is divided as follows: we present a brief review of the Implicit Regularization (IReg) scheme in the next section, we then 
apply it to calculate the Ward identities of the Schwinger model and of the Adler-Bardeen-Bell-Jackiw anomaly in sections \ref{Schwinger}
and \ref{ABJ}, respectively. In section \ref{s5} we show that gauge symmetry in chiral abelian gauge theories is fulfilled if we require Momentum Routing Invariance (MRI)
and vice-versa. Finally, we present our conclusions in section \ref{conclusion}.

\section{Implicit Regularization in a nutshell}
\label{IReg}

We apply the IReg framework \cite{IR} to treat the integrals which appear in the amplitudes of the next sections. Let 
us make a brief review of the method in four dimensions. In this scheme, we assume that the integrals are regularized by an implicit 
regulator $\Lambda$ just to justify algebraic operations within the integrands. We then use the following identity 
\begin{equation}
\int_k\frac{1}{(k+p)^2-m^2} =\int_k\frac{1}{k^2-m^2}
 -\int_k\frac{(p^2+2p\cdot k)}{(k^2-m^2)[(k+p)^2-m^2]},
\label{2.1}
\end{equation}
where $\int_k\equiv\int^\Lambda\frac{d^4 k}{(2\pi)^4}$, to separate basic divergent integrals (BDI's) from the finite part. These BDI's 
are defined as follows
\begin{equation}
I^{\mu_1 \cdots \mu_{2n}}_{log}(m^2)\equiv \int_k \frac{k^{\mu_1}\cdots k^{\mu_{2n}}}{(k^2-m^2)^{2+n}}
\end{equation}
and
\begin{equation}
I^{\mu_1 \cdots \mu_{2n}}_{quad}(m^2)\equiv \int_k \frac{k^{\mu_1}\cdots k^{\mu_{2n}}}{(k^2-m^2)^{1+n}}.
\end{equation}

The basic divergences with Lorentz indexes can be judiciously combined as differences between integrals with the same superficial degree 
of divergence, according to the equations below, which define surface terms  \footnote{The Lorentz indexes between brackets stand for 
permutations, i.e. $A^{\{\alpha_1\cdots\alpha_n\}}B^{\{\beta_1\cdots\beta_n\}}=A^{\alpha_1\cdots\alpha_{n}}B^{\beta_1\cdots\beta_n}$ 
+ sum over permutations between the two sets of indexes $\alpha_1\cdots\alpha_{n}$ and $\beta_1\cdots\beta_n$}:
\begin{eqnarray}
\Upsilon^{\mu \nu}_{2w}=  g^{\mu \nu}I_{2w}(m^2)-2(2-w)I^{\mu \nu}_{2w}(m^2) 
\equiv \upsilon_{2w}g^{\mu \nu},
\label{dif1}\\
\nonumber\\
\Xi^{\mu \nu \alpha \beta}_{2w}=  g^{\{ \mu \nu} g^{ \alpha \beta \}}I_{2w}(m^2)
 -4(3-w)(2-w)I^{\mu \nu \alpha \beta }_{2w}(m^2)\equiv\nonumber\\
\equiv  \xi_{2w}(g^{\mu \nu} g^{\alpha \beta}+g^{\mu \alpha} g^{\nu \beta}+g^{\mu \beta} g^{\nu \alpha}).
\label{dif2}
\end{eqnarray} 

In the expressions above, $2w$ is the degree of divergence of the integrals and  for the sake of brevity, we substitute the subscripts 
$log$ and $quad$ by $0$ and $2$, respectively. Surface terms can be conveniently written as integrals of total 
derivatives, namely
\begin{eqnarray}
\upsilon_{2w}g^{\mu \nu}= \int_k\frac{\partial}{\partial k_{\nu}}\frac{k^{\mu}}{(k^2-m^2)^{2-w}},
\label{ts1}
\end{eqnarray}
\begin{eqnarray}
(\xi_{2w}-v_{2w})(g^{\mu \nu} g^{\alpha \beta}+g^{\mu \alpha} g^{\nu \beta}+g^{\mu \beta} g^{\nu \alpha})= \int_k\frac{\partial}{\partial 
k_{\nu}}\frac{2(2-w)k^{\mu} k^{\alpha} k^{\beta}}{(k^2-m^2)^{3-w}}.
\label{ts2}
\end{eqnarray}

We see that equations (\ref{dif1})-(\ref{dif2}) are undetermined because they are differences between divergent quantities  with the same degree of divergence. Each 
regularization scheme gives a different value for these terms. However, as Physics should not depend on the scheme applied, we leave 
these terms to be arbitrary until the end of the calculation, fixing them by symmetry constraints or phenomenology, when it applies \cite
{JackiwFU}. Similar considerations were also envisaged by the author of \cite{Jacquot:1999hj}.

Of course the same idea can be used at any dimension of space-time. In (1+1)- dimensions, for instance, a basic logarithmic divergent 
integral would be like $\int \frac{d^2 k}{(2\pi)^2}\frac{1}{k^2-m^2}$ and a logarithmic surface term would be like $g^{\mu 
\nu}\upsilon_0=g^{\mu \nu}\int \frac{d^2 k}{(2\pi)^2} \frac{1}{(k^2-m^2)}-2\int \frac{d^2 
k}{(2\pi)^2}\frac{k^{\mu}k^{\nu}}{(k^2-m^2)^{2}}$.

\section{Chiral Schwinger model}
\label{Schwinger}

In this section we focus on the chiral Schwinger model in order to have a simple example on regularization dependent identities regarding $\gamma_5$. We also discuss the anomaly appearing in this model in our framework, showing how it can be democratically displayed between the axial and vectorial Ward identities.

The Chiral Schwinger model was first considered in \cite{Jackiw}. It was shown that although the theory is unitary, it is not gauge invariant at the quantum level. We are going to recover this result at the end of this section (when we consider the complete one loop two-point function for the gauge boson). However, in order to make contact to the ABJ anomaly case, which will be addressed in the next section, we consider first an effective case in which the two-point function contains a definite axial and vectorial vertex. We show then that it is only possible to preserve one of the Ward identities at a time, either the vectorial or the axial one.

The model we are going to study is given by the following induced action \cite{JackiwFU}
\be
\Gamma_{CS}(A)=-i\ln \det (i\slashed{\partial}-e(1+\gamma_{5})\slashed{A}).
\ee

We focus on the two-point function of the photon with a vectorial and a chiral vertex. It's amplitude is given by
\be
\Pi_{\mu\nu}=-ie^{2}Tr\int_{q}\gamma_{\mu}\frac{1}{\slashed{q}-\slashed{p}}\gamma_{\nu}\gamma_{5}\frac{1}{\slashed{q}},
\ee
where $\int_{q}$ stands for $\int \frac{d^{2}q}{(2\pi)^{2}}$.
%
 
At this point some regularization must be adopted in order to deal properly with this integral. We will choose the IReg (IReg) framework which, as stated in 
the introduction, preserves the space time dimension of the underlying theory at all times. At first view, one could make the following statement: since the 
algebra of the $\gamma_{5}$ is unambiguous only in integer dimensions, any identity involving such object should remain true in a dimension preserving 
method such as IReg. Surprisingly enough, this statement reveals to be false, as we are going to exemplify. In fact, even though identities involving 
$\gamma_{5}$, in particular $\{\gamma_{\mu},\gamma_{5}\}=0$, are true in integer dimensions, they may be false when inside a divergent integral. 

We return to our example. The first prescription for dealing with the $\gamma_{5}$ will be:
\be
\gamma_{\nu}\gamma_{5}=\epsilon_{\nu\theta}\gamma^{\theta}.
\ee   
After this step, the evaluation of the integral using the IReg rules is straightforward furnishing
\be
\Pi_{\mu\nu}=-2ie^{2}\epsilon_{\nu\theta}\left[\frac{(\delta_{\mu}^{\theta}p^{2}-p_{\mu}p^{\theta})(-2b)}{p^{2}}-\delta_{\mu}^{\theta}\upsilon_{0}\right],
\label{first}
\ee
where $b=i/4\pi$, and $\upsilon_{0}$ is a surface term which is ambiguous. We list all integrals found in this amplitude in the 
appendix.

It is instructive to compute the two Ward identities, the vectorial and axial one respectively:
\begin{align}
p^{\mu}\Pi_{\mu\nu}=2ie^{2}\epsilon_{\nu\theta}p^{\theta}\upsilon_{0},\nonumber\\
p^{\nu}\Pi_{\mu\nu}=-2ie^{2}\epsilon_{\mu\theta}p^{\theta}(2b+\upsilon_{0}).
\label{WIS}
\end{align}

We notice the appearance of an arbitrariness parametrized as a surface term which allows a democratic view of the anomaly. In fact, in the IReg framework, 
symmetries or phenomenology should be the guide to fix ambiguities instead of the regularization method by itself. In this sense, by suitable choices for 
the $\upsilon_{0}$ one can preserve one of the identities ($\upsilon_{0}=0$ preserves the vectorial one while $\upsilon_{0}=-2b$ preserves the axial one) or 
even distribute the anomaly between both identities (choosing $\upsilon_{0}=-b$).

We use now the identity
\be
\{\gamma_{\mu},\gamma_{5}\}=0,
\label{anti}
\ee
which is expected to yield the same result. Thus
\be
\Pi_{\mu\nu}=ie^{2}Tr\int_{q}\gamma_{\mu}\frac{1}{\slashed{q}-\slashed{p}}\gamma_{5}\gamma_{\nu}\frac{1}{\slashed{q}}.
\ee

As before, some prescription is still needed to deal with the $\gamma_{5}$, which we choose to be
\be
\frac{1}{\slashed{q}-\slashed{p}}\gamma_{5}=\frac{(q-p)^{\alpha}}{(q-p)^{2}}\gamma_{\alpha}\gamma_{5}=\frac{(q-p)^{\alpha}}{(q-p)^{2}}\epsilon_{\alpha\theta}\gamma^{\theta}.
\label{anti2}
\ee

The evaluation of $\Pi_{\mu\nu}$ is straightforward
\be
\Pi_{\mu\nu}=2ie^{2}\frac{b}{p^{2}}\left(p^{\alpha}\epsilon_{\alpha\nu}p_{\mu}+p^{\alpha}\epsilon_{\alpha\mu}p_{\nu}\right),
\label{second}
\ee 
which differs from the result in equation (\ref{first}).  For the Ward identities one has
\begin{align}
p^{\mu}\Pi_{\mu\nu}=-2ie^{2}\epsilon_{\nu\theta}p^{\theta}b,\nonumber\\
p^{\nu}\Pi_{\mu\nu}=-2ie^{2}\epsilon_{\mu\theta}p^{\theta}b.
\end{align}

The first point to be noticed is the disappearance of the surface term. This allows us to conjecture that, by using the prescriptions stated in equations (\ref{anti}) and (\ref{anti2}), one is implicitly evaluating some of the divergent integrals. For instance, by choosing $\upsilon_{0}=-b$ in the Ward identities of the first case one obtains the equations above.

Therefore, it seems that the way one chooses to deal with dimension specific objects (such as $\gamma_{5}$) inside divergent integrals is ambiguous. The question now is which prescription should one rely on. From our point of view, the computation should be performed in the most democratic possible way which means traces involving $\gamma_{5}$ must contain all possible Lorentz structures available. Therefore, one should adopt the following symmetric trace
\begin{align}
Tr(\gamma_{\sigma}\gamma_{\mu}\gamma_{\alpha}\gamma_{\nu}\gamma_{5})&=2\left(-\epsilon_{\sigma\nu}g_{\alpha\mu}+
\epsilon_{\mu\nu}g_{\alpha\sigma}-\epsilon_{\alpha\nu}g_{\sigma\mu}+\epsilon_{\sigma\alpha}g_{\mu\nu}
-\epsilon_{\mu\alpha}g_{\sigma\nu}-\epsilon_{\sigma\mu}g_{\alpha\nu}\right),
\label{traco}
\end{align}
\noindent
which is obtained replacing $\gamma_{5}$ by its definition on a two-dimensional space ($\gamma_{5} = \epsilon_{\mu\rho}\gamma^{\mu}\gamma^{\rho}/2$).

Replacing this identity in the amplitude and using the prescription $\gamma_{a}\gamma_{5}=\epsilon_{a\theta}\gamma^{\theta}$ one finally obtains
\be
\Pi_{\mu\nu}=-2ie^{2}\epsilon_{\nu\theta}\left[\frac{(\delta_{\mu}^{\theta}p^{2}-p_{\mu}p^{\theta})(-2b)}{p^{2}}-\delta_{\mu}^{\theta}\upsilon_{0}\right].
\label{first2}
\ee

It should be noticed that this is the result of eq. (\ref{first}) and the explanation is the following. As can be seen from the equation above, even though the trace contains all possible structures for the Levi-Civita tensor, in the end only $\epsilon_{\nu\theta}$ appears. In the first case considered, this was our choice from the beginning through the prescription $\gamma_{\nu}\gamma_{5}=\epsilon_{\nu\theta}\gamma^{\theta}$. Therefore, it is no surprise that the two results coincide. 

Another interesting point is the connection between surface terms and shifts in the internal momenta. This aspect is more evident in the Ward identities which we quote below
\begin{align}
p^{\mu}\Pi_{\mu\nu}&=2ie^{2}\epsilon_{\nu\theta}p^{\theta}\upsilon_{0},\nonumber\\
p^{\nu}\Pi_{\mu\nu}&=-2ie^{2}\epsilon_{\mu\theta}p^{\theta}(2b+\upsilon_{0}).
\end{align}

As has been shown, this identities were obtained by computing $\Pi_{\mu\nu}$ and afterward contracting with the external momentum $p^{\mu}$. One might wonder whether performing the contraction before would pose any problem. As can be easily seen, one obtains for the vectorial Ward identity 
\begin{align}
p^{\mu}\Pi_{\mu\nu}=-ie^{2}\left[Tr\int_{q}\frac{1}{\slashed{q}-\slashed{p}}\gamma_{\nu}\gamma_{5}-Tr\int_{q}\frac{1}{\slashed{q}}\gamma_{\nu}\gamma_{5}\right],
\end{align}
which means that, if shifts in the internal momentum were naively allowed, one would automatically obtain the conservation of the vectorial Ward identity. This is the result of Dimensional Regularization since in this method shifts are always allowed. From our perspective, this aspect should be taken with care. As presented in \cite{Cherchiglia}, MRI (which from a mathematical point of view reveals itself to be equivalent as performing shifts in the internal momentum) is always connected with the appearance of surface terms and can only be satisfied if the latter are always set to zero. Therefore, using the results of \cite{Cherchiglia} one obtains
\begin{align}
p^{\mu}\Pi_{\mu\nu}=2ie^{2}\epsilon_{\nu\theta}p^{\theta}\upsilon_{0},
\end{align}
in agreement with our previous result. Therefore, for the vectorial Ward identity, contracting with the external momentum $p$ before or after evaluating the amplitude, is innocuous. 

This is however not anymore the case for the axial Ward identity, as can be easily seen, after contracting the amplitude with the external momentum $p^{\nu}$ and using that $\slashed{p}=\slashed{p}-\slashed{q}+\slashed{q}$. One obtains
\begin{align}
p^{\nu}\Pi_{\mu\nu}=-ie^{2}\left[Tr\int_{q}\gamma_{\mu}\frac{1}{\slashed{q}-\slashed{p}}\slashed{q}\gamma_{5}\frac{1}{\slashed{q}}-Tr\int_{q}\gamma_{\mu}\gamma_{5}\frac{1}{\slashed{q}}\right].
\end{align}
\noindent
Naively one could use $\{\gamma_{\mu},\gamma_{5}\}=0$ which would furnish
\begin{align}
p^{\nu}\Pi_{\mu\nu}=&-ie^{2}\left[Tr\int_{q}\gamma_{\mu}\gamma_{5}\frac{1}{\slashed{q}-\slashed{p}}-Tr\int_{q}\gamma_{\mu}\gamma_{5}\frac{1}{\slashed{q}}\right],\nonumber\\
=&2ie^{2}\epsilon_{\mu\theta}p^{\theta}\upsilon_{0},
\end{align}
in contrast with our previous result for the axial Ward identity, eq. (\ref{WIS}). It is, however, compatible with the result we obtained when using $\{\gamma_{\mu},\gamma_{5}\}=0$, computing the amplitude and \textit{then} contracting with $p^{\nu}$, if one chooses $\upsilon_{0}=-b$. Therefore, once again we show that allowing $\{\gamma_{\mu},\gamma_{5}\}=0$ to be applied inside divergent integrals is an ambiguous operation. In the latter example this feature is even more dramatic since it does not define the value of the anomaly, all one can infer is that the arbitrariness is equally distributed in the two Ward identities.  

We discuss now how the last calculation should be performed, in order to define the value of the anomaly and still respect democracy. As before the problem resides in the naive application of $\{\gamma_{\mu},\gamma_{5}\}=0$. To avoid this, one can just use the symmetric formula for the trace (eq. \ref{traco}) as we did before. An equivalent approach, more closely related to the previous analysis, is to first use the identity $\{\gamma_{\mu},\gamma_{\nu}\}=2g_{\mu\nu}$ to rewrite the amplitude and then use the definition of $\gamma_{5}$ to evaluate the integrals. Explicitly,
\begin{align}
p^{\nu}\Pi_{\mu\nu}=&\underbrace{-ie^{2}\left[-Tr\int_{q}\frac{1}{\slashed{q}-\slashed{p}}\gamma_{\mu}\gamma_{5}-Tr\int_{q}\gamma_{\mu}\gamma_{5}\frac{1}{\slashed{q}}\right]}_{-2ie^{2}\epsilon_{\mu\theta}p^{\theta}\upsilon_{0}}+\nonumber\\&\underbrace{-ie^{2}\left[2Tr\int_{q}\frac{(q-p)_{\mu}}{(q-p)^2}\slashed{q}\gamma_{5}\frac{1}{\slashed{q}}-2Tr\int_{q}\frac{q_{\mu}}{\slashed{q}-\slashed{p}}\gamma_{5}\frac{1}{\slashed{q}}+2Tr\int_{q}\frac{q^{\sigma}(q-p)_{\sigma}}{(q-p)^2}\gamma_{\mu}\gamma_{5}\frac{1}{\slashed{q}}\right]}_{-4ie^{2}\epsilon_{\mu\theta}p^{\theta}b},\nonumber\\
=&-2ie^{2}\epsilon_{\mu\theta}p^{\theta}(2b+\upsilon_{0}),
\end{align}
which is the result obtained before, respecting the democracy the calculation should retain \textit{and} defining the precise value of the anomaly.

In summary, although $\{\gamma_{\mu},\gamma_{5}\}=0$ is well-defined in integer dimensions, this furnishes ambiguous results inside divergent integrals 
independently if one uses dimensional extensions (as in Dimensional Regularization) or stays in the physical dimension of the theory (as we did).
Although the usual procedure is to extend the algebra of $\gamma_5$ matrices  \cite{Tsai1,Tsai2,Ferrari1,Ferrari2,Ferrari3,Bonneau,Thompson,Baikov,Zralek} to $D$ dimensions, from our perspective, this 
operation should be avoided and, whenever dimension specific objects such as $\gamma_{5}$ appear, one should use the most democratic expression available. 
In our case, we had to adopt a democratic expression for the trace of four $\gamma$ matrices and $\gamma_{5}$ in the sense that all possible Lorentz 
structures available were present. 

To conclude this section, we would like to comment on the computation of the complete two-point function in the Chiral Schwinger Model. In this case, both vertices will contain a factor $\gamma_{\mu}(1+\gamma_5)$ and one needs to compute the trace not only with one but two $\gamma_5$ matrices. Relying on the findings of the present section, we propose that the most general form for the trace should be used which amounts to replacing both $\gamma_5$ by their definition on a two-dimensional space ($\gamma_{5} = \epsilon_{\mu\rho}\gamma^{\mu}\gamma^{\rho}/2$). The net result will be 
\begin{align}
\Pi_{\mu\nu}^{\mbox{full}}&=-ie^{2}Tr\int_{q}\gamma_{\mu}(1+\gamma_{5})\frac{1}{\slashed{q}-\slashed{p}}\gamma_{\nu}(1+\gamma_{5})\frac{1}{\slashed{q}},\nonumber\\
&=-\frac{e^{2}}{\pi}\left[2\left(g_{\mu\nu}-\frac{p_{\mu}p_{\nu}}{p^{2}}\right)-\frac{p_{\mu}\epsilon_{\nu\theta}p^{\theta}}{p^{2}}-\frac{p_{\nu}\epsilon_{\mu\theta}p^{\theta}}{p^{2}}\right]+4ie^{2}g_{\mu\nu}\upsilon_{0}.
\end{align}
\noindent 
where the contributions containing just one $\gamma_{5}$ are given by eq. (\ref{first}) while the ones proportional to none (diagram VV) or two $\gamma_{5}$ (diagram AA) give the same result namely
\be
\Pi_{\mu\nu}^{\mbox{VV}}=\Pi_{\mu\nu}^{\mbox{AA}}=-2ie^{2}\left[\frac{(g_{\mu\nu}p^{2}-p_{\mu}p^{\nu})(-2b)}{p^{2}}-g_{\mu\nu}\upsilon_{0}\right].
\ee

A characteristic of the model (already commented by some of us in \cite{BaetaScarpelli:2000zs}) is the impossibility to obtain a gauge invariant result for any value of the arbitrariness $\upsilon_{0}$. Nevertheless, the appearance of the surface term is vital since only for positive values of $\lambda\equiv -4i\pi\upsilon_{0}$ one obtains a sensible (unitary) theory \cite{Jackiw} which contains a radiatively induced massive gauge boson  with mass
\be
m^2=\frac{e^2}{\pi}\frac{(\lambda+1)^2}{\lambda},
\ee
in agreement with \cite{Jackiw}.

\section{Revisiting the Adler-Bardeen-Bell-Jackiw anomaly}
\label{ABJ}

Since its discovery \cite{{BellJackiw},{Adler}}, the ABJ anomaly has been calculated by several approaches \cite{Elias,YU,Tsai1,Ferrari2,Cynolter,Perez}. Including the recent 
rightmost position approach \cite{Tsai1}, prescriptions to deal with $\gamma_5$ matrices of this amplitude are sought after, which allow 
Dimensional Regularization to preserve gauge symmetry. An overview on the various regularization
schemes applied in the diagrammatic anomaly computation can be found in \cite{Bertlmann}. There 
are also derivations obtained by the path integral measure transformation 
\cite{Fujikawa} and by differential geometry \cite{Zee}. The usual view on the diagrammatic anomaly derivation is that this anomaly 
occurs due to the momentum routing breaking in the internal lines. The momenta of those internal lines must assume specific values to fulfill 
the Ward identity we want to preserve.

In this section we derive the ABJ anomaly by means of IReg. This approach relies in the democratic display of the
Ward identities. All the symmetry breaking information  is contained in the surface term whose value is determined by the Ward 
identity we want to preserve. That is what has been done in section \ref
{Schwinger} as seen in equations (\ref{WIS}). In the neutral pion decay in two photons, the vector Ward identities must be preserved and, consequently, the axial one is violated \cite{{BellJackiw},{Adler}}. On the other
hand, in the Standard Model the chiral coupling with gauge fields refers to fermion-number conservation and the axial identity must be 
enforced \cite{tHooft}.

A further advantage of this approach is that it leads to a momentum routing independent result. As we are going to show, we 
perform this calculation with arbitrary routings of the internal momenta. Those arbitrary routings multiply the arbitrary surface term
in the final result. From the physical point of view, it is more appealing to choose a value for the surface term instead of choosing
the routing, since the former is a difference of divergent integrals whose result is unknown, while  the routing of momenta should be kept  unconstrained  as long as momentum conservation at the vertices is fulfilled.

Before we proceed, let us comment on objects like $Tr[\gamma ^{\mu }\gamma ^{\beta }\gamma ^{\nu }\gamma ^{\xi }\gamma ^{\alpha 
}\gamma ^{\lambda }\gamma ^5]$ that are found in this amplitude. It is possible to use the following identity to reduce the number of Dirac
$\gamma$ matrices
\be
\gamma ^{\mu }\gamma ^{\beta }\gamma ^{\nu }= g^{\mu \beta} \gamma^{\nu}+g^{\nu \beta} \gamma^{\mu}-g^{\mu \nu} \gamma^{\beta}-i 
\epsilon^{\mu \beta \nu \rho}\gamma_{\rho}\gamma_5.
\label{gamma}
\ee

Using equation (\ref{gamma}), $Tr[\gamma ^{\mu }\gamma ^{\beta }\gamma ^{\nu }\gamma ^{\xi }\gamma ^5]=4i\epsilon^{\mu \beta \nu \xi}$ 
and $\gamma_5\gamma^{\rho}\gamma_5=-\gamma^{\rho}$ leads to the following result
\be
Tr[\gamma ^{\mu }\gamma ^{\beta }\gamma ^{\nu }\gamma ^{\xi }\gamma ^{\alpha}\gamma ^{\lambda }\gamma ^5]=4i(g^{\beta \mu}\epsilon^{\nu 
\xi \alpha \lambda}+g^{\beta \nu}\epsilon^{\mu \xi \alpha \lambda}-g^{\mu \nu}\epsilon^{\beta \xi \alpha \lambda}
-g^{\lambda \alpha}\epsilon^{\mu \beta \nu \xi}+g^{\xi \lambda}\epsilon^{\mu \beta \nu \alpha}-g^{\xi \alpha}\epsilon^{\mu \beta \nu 
\lambda}).
\label{trace}
\ee

However, it is completely arbitrary which three $\gamma$ matrices are taken to apply equation (\ref{gamma}). A different choice would 
give the result of equation (\ref{trace}) with Lorentz indexes permuted. Furthermore, equation (\ref{anti}) should be avoided inside 
a divergent integral as we could see in section \ref{Schwinger}, since this operation seems to fix a value for the surface term. This point
of view is also shared by \cite{Cynolter}. Therefore, we adopt a four dimensional version of equation (\ref{traco}), which contains all 
possible Lorentz structures available. This equation reads
\begin{align}
\frac{-i}{4}Tr[\gamma ^{\mu }\gamma ^{\nu }\gamma ^{\alpha }\gamma ^{\beta }\gamma ^{\gamma }\gamma ^{\delta }\gamma ^5]&=-g^{\alpha 
\beta } \epsilon ^{\gamma \delta \mu \nu }+g^{\alpha \gamma } \epsilon ^{\beta \delta \mu \nu }-g^{\alpha \delta } \epsilon ^{\beta 
\gamma \mu \nu }-g^{\alpha \mu } \epsilon ^{\beta \gamma \delta \nu }+g^{\alpha \nu } \epsilon ^{\beta \gamma \delta \mu }-g^{\beta 
\gamma } \epsilon ^{\alpha \delta \mu \nu }+\nonumber\\&+g^{\beta \delta } \epsilon ^{\alpha \gamma \mu \nu }+g^{\beta \mu } \epsilon ^{
\alpha \gamma \delta \nu }-g^{\beta \nu } \epsilon ^{\alpha \gamma \delta \mu }-g^{\gamma \delta } \epsilon ^{\alpha \beta \mu \nu }-g^{
\gamma \mu } \epsilon ^{\alpha \beta \delta \nu }+g^{\gamma \nu } \epsilon ^{\alpha \beta \delta \mu }+\nonumber\\&+g^{\delta \mu } 
\epsilon ^{\alpha \beta \gamma \nu }-g^{\delta \nu } \epsilon ^{\alpha \beta \gamma \mu }-g^{\mu \nu } \epsilon ^{\alpha \beta \gamma 
\delta },
\label{traco2}
\end{align}
which can be obtained replacing $\gamma_5$ by its definition at four dimensions, 
$\gamma_5=\frac{i}{4!}\epsilon^{\mu\nu\alpha\beta}\gamma_{\mu}\gamma_{\nu}\gamma_{\alpha}\gamma_{\beta}$.

We have to use this unambiguous result in the Feynman amplitude for the triangle diagram whenever we find a trace of six $\gamma$ 
matrices with a $\gamma_5$ matrix. Equation (\ref{traco2}) has already been used previously in other works \cite{{Cynolter},{Perez},{Wu}}.
\begin{figure}[!h]
\includegraphics[trim=0mm 100mm 0mm 100mm,scale=0.5]{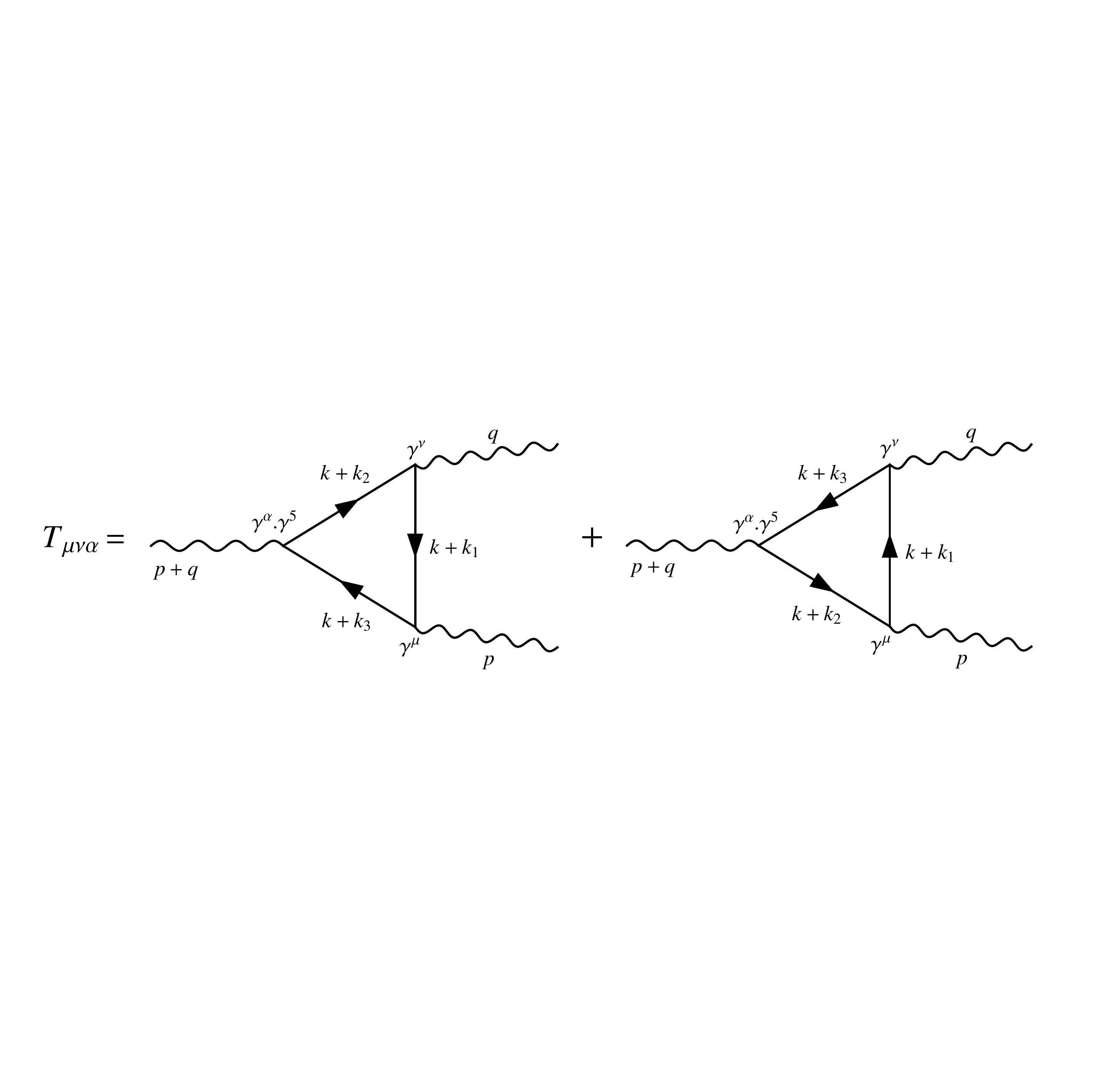}
\caption{Triangle diagrams which contribute to the ABJ anomaly. We label the internal lines with arbitrary momentum routing.}
\label{triangle}
\end{figure}
The amplitude of the Feynman diagrams of figure \ref{triangle} is given by
\be
T_{\mu\nu\alpha}=-i\int_k Tr\left[\gamma_{\mu}\frac{i}{\slashed{k}+\slashed{k}_1-m}\gamma_{\nu}\frac{i}{\slashed{k}+\slashed{k}_2-m}
\gamma_{\alpha}\gamma_5 \frac{i}{\slashed{k}+\slashed{k}_3-m}\right]+(\mu \leftrightarrow \nu, p\leftrightarrow q).
\label{AVV}
\ee
where the arbitrary routing $k_i$ obeys the following relations due to energy-momentum conservation at each vertex
\begin{align}
k_2-k_3=& p+q,\nonumber\\
k_1-k_3=& p,\nonumber\\
k_2-k_1=& q.
\label{routing}
\end{align}

Equations (\ref{routing}) allow us to parametrize the routing  $k_i$ as
\begin{align}
k_1=& \alpha p+(\beta-1) q,\nonumber\\
k_2=& \alpha p+\beta q,\nonumber\\
k_3=& (\alpha-1) p+(\beta-1) q,
\label{routing2}
\end{align}
where $\alpha$ and $\beta$ are arbitrary real numbers which map the freedom we have in choosing the routing of internal lines, i. e. we 
may add any combination of $q$ and $p$ in each internal line as long we respect the momentum conservation given by equations (\ref{routing}).
Equations (\ref{routing}) and (\ref{routing2}) for the other diagram are obtained by changing $ p\leftrightarrow q$.

After taking the trace using equation (\ref{traco2}), we apply the IReg scheme in order to regularize the integrals coming from equation (\ref{AVV}). The result is 
\be
T_{\mu\nu\alpha}= 4i\upsilon_0 (\alpha-\beta-1)\epsilon_{\mu \nu \alpha \beta}(q-p)^{\beta}+T^{finite}_{\mu\nu\alpha},
\label{resultado}
\ee
where $\upsilon_0$ is a surface term defined in section \ref{IReg} and $T^{finite}_{\mu\nu\alpha}$ is the finite part of the amplitude 
whose evaluation we perform in the appendix.

We then apply the respective external momentum in equation (\ref{resultado}) in order to obtain the Ward identities:
\begin{align}
&p_{\mu}T^{\mu \nu \alpha}=-4i\upsilon_0 (\alpha-\beta-1)\epsilon^{\alpha \nu \beta \lambda}p_{\beta}q_{\lambda},\nonumber\\
&q_{\nu}T^{\mu \nu \alpha}=4i\upsilon_0 (\alpha-\beta-1)\epsilon^{\alpha \mu \beta \lambda}p_{\beta}q_{\lambda},\nonumber\\
&(p+q)_{\alpha}T^{\mu \nu \alpha}=2m T_5^{\mu \nu}+8i\upsilon_0 (\alpha-\beta-1)\epsilon^{\mu \nu \beta \lambda}p_{\beta}q_{\lambda}-\frac
{1}{2\pi^2}\epsilon^{\mu \nu \beta \lambda}p_{\beta}q_{\lambda},
\label{WI}
\end{align}
where $T_5^{\mu \nu}$ is the usual vector-vector-pseudoscalar triangle.

The number $\upsilon_0 (\alpha-\beta-1)$ is arbitrary since $\upsilon_0$ is a difference of
two infinities and $\alpha$ and $\beta$ are any real numbers that we have freedom in choosing as long as equations (\ref{routing}) 
representing the energy-momentum conservation hold. We can parametrize this arbitrariness in a single parameter $a$ redefining 
$4i\upsilon_0(\alpha-\beta-1)\equiv\frac{1}{4\pi^2}(1+a)$. Equations (\ref{WI}) reads
\begin{align}
&p_{\mu}T^{\mu \nu \alpha}=-\frac{1}{4\pi^2}(1+a)\epsilon^{\alpha \nu \beta \lambda}p_{\beta}q_{\lambda},\nonumber\\
&q_{\nu}T^{\mu \nu \alpha}=\frac{1}{4\pi^2}(1+a)\epsilon^{\alpha \mu \beta \lambda}p_{\beta}q_{\lambda},\nonumber\\
&(p+q)_{\alpha}T^{\mu \nu \alpha}=2m T_5^{\mu \nu}+\frac{1}{2\pi^2}a\epsilon^{\mu \nu \beta \lambda}p_{\beta}q_{\lambda}.
\label{WI22}
\end{align}
\noindent
From now on, we will focus only in the massless theory since we would like to discuss just the quantum symmetry breaking term, namely the ABJ anomaly. 

As in the Schwinger model, we can see a democracy displayed in the Ward identities (\ref{WI22}). If gauge invariance is to be preserved, one takes $a=-1$ and automatically the axial identity is violated by a quantity equal to $-\frac{1}{2\pi^2}\epsilon^{\mu \nu \beta \lambda}p_{\beta}q_{\lambda}$. On the other hand,
if chiral symmetry is maintained at the quantum level, we choose $a=0$, and the vectorial identities are violated. The choice
$a=-1$ sets the surface term to zero. This has already been observed by some of us in \cite{Cherchiglia}. It was proved that setting
surface terms to zero assures gauge invariance and momentum routing invariance of abelian gauge theories. However, the results (\ref{WI22}) 
of the chiral anomaly are the opposite of those obtained by some of us in a previous work \cite{Scarpelli}. This is because traces like
$Tr[\gamma ^{\mu }\gamma ^{\nu }\gamma ^{\alpha }\gamma ^{\beta }\gamma ^{\gamma }\gamma ^{\delta }\gamma ^5]$  were derived considering 
the anti-commutation relation (\ref{anti}), i. e. the result of equation (\ref{trace}), which as in the previous example may fix a 
finite value for the surface term $\upsilon_0$. 

Furthermore, in \cite{Scarpelli} the vectorial Ward identities were satisfied when the surface term
assumed a non-null value such that it canceled with the finite part in order to preserve gauge symmetry. Therefore, it was thought
that the anomaly was due to the breaking of the MRI. That conclusion was supported by \cite{Cherchiglia} where it was shown that making 
surface terms null is a necessary and sufficient condition to assure gauge and momentum routing invariance of abelian theories. The 
former invariance is a consequence of the latter and conversely.


In the literature about diagrammatic computation of anomalies it is consensus that the breaking of the MRI, i.e. 
the necessity to choose an internal routing, is required to implement the conservation of the vector Ward identities. However we have 
shown that taking advantage of IReg supplemented by symmetrization of the trace, arbitrary routing is conform with gauge invariance, and 
so the result is MRI. The reason is that any such arbitrariness is always accompanied with a surface term, which is set to zero on gauge 
invariance grounds. Once this is accepted, and following the same line of reasoning, one can infer that MRI is also preserved  for the 
case that the axial Ward identity is  verified, since arbitrary routing will finally also be absorbed in the choice of the surface term.
Although these affirmations may be taken to be just semantics, it is our opinion that they shed light on the interpretation of MRI in 
diagrammatic calculations and the role of surface terms. As will be discussed in the next section, some processes involve only surface 
terms and no arbitrary routing dependence, thus they are manifestly MRI, which will allow to understand the role of the surface term 
alone.

\section{Gauge and Momentum Routing Invariance in Chiral Abelian Gauge Theories}
\label{s5}

In this section we study the connection between momentum routing invariance (MRI) and vectorial gauge symmetry in the case of effective Chiral Abelian Gauge Theories to arbitrary order in perturbation theory. We will adopt a diagrammatic point of view and show how the proof of gauge invariance in Abelian Gauge Theories already done in \cite{Cherchiglia} can be extended to the present case. Before we proceed, it should be emphasized that what we intend to prove is that MRI is still connected with \textbf{vectorial} gauge symmetry even in the case in which an axial coupling between fermions is allowed. Therefore, for simplicity, we just consider that one of the vertices of the diagrams used in the diagrammatic proof is an axial one, instead of considering the general case with more axial vertices.

\begin{figure}[!h]
\includegraphics[trim=50mm 90mm 0mm 10mm,scale=0.5]{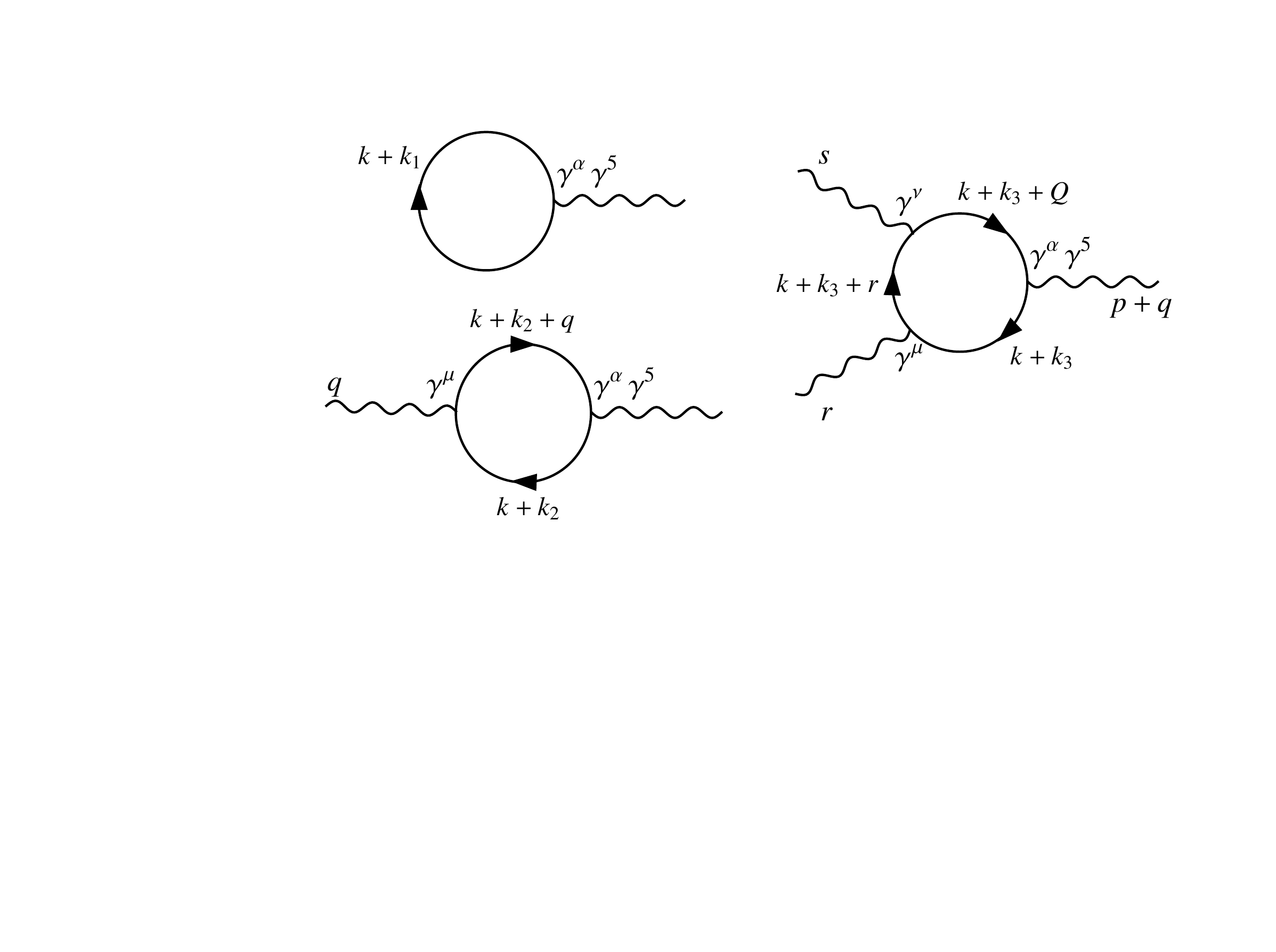}
\caption{Diagrams upon which the diagrammatic proof of gauge invariance is constructed.}
\label{FIGURA1}
\end{figure}

\begin{figure}[!h]
\includegraphics[trim=20mm 60mm 80mm 60mm,scale=0.4]{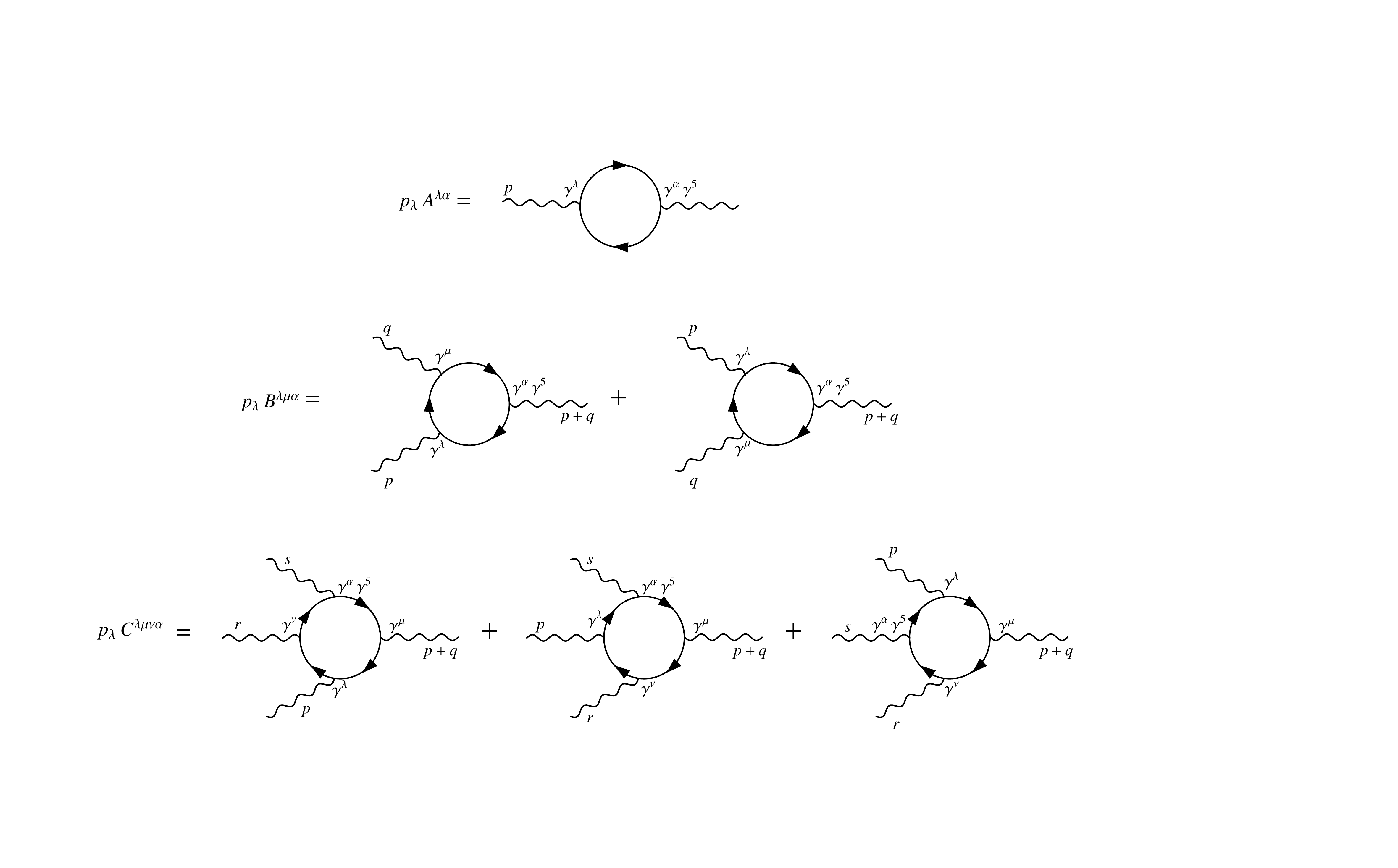}
\caption{Pictorial representation of the Ward identities.}
\label{figd}
\end{figure}

As explained in \cite{Cherchiglia}, the starting point for the diagrammatic proof of gauge invariance are the diagrams depicted in fig. \ref{FIGURA1} in which the external momenta $p$ is inserted in all possible ways furnishing a pictorial representation for Ward identities \cite{Peskin} as depicted in figure \ref{figd}. 

%
Explicitly one obtains:
\begin{align}
&p_{\lambda}A^{\lambda\alpha}=\int_k Tr \left[\frac{1}{\slashed{k}+\slashed{k}_1-m}\slashed{p}
\frac{1}{\slashed{k}+\slashed{k}_1+\slashed{p}-m}\gamma^{\alpha}\gamma^{5}\right]
\label{bolha}\\
p_{\lambda}B^{\lambda\mu\alpha}=&\int_k Tr \left[\frac{1}{\slashed{k}+\slashed{k}_2-m}\slashed{p}\frac{1}{\slashed{k}+\slashed{k}_2+\slashed{p}-m}\gamma^{\mu}\frac{1}{\slashed{k}+\slashed{k}_2+\slashed{p}+\slashed{q}-m}\gamma^{\alpha}\gamma^{5}\right]+\nonumber\\
&+\int_k Tr \left[\frac{1}{\slashed{k}+\slashed{k}_2-m}\gamma^{\mu}\frac{1}{\slashed{k}+\slashed{k}_2+\slashed{q}-m}\slashed{p}\frac{1}{\slashed{k}+\slashed{k}_2+\slashed{p}+\slashed{q}-m}\gamma^{\alpha}\gamma^{5}\right]
\label{bolha2}
\end{align}
\begin{align}
p_{\lambda}C^{\lambda\mu\nu\alpha}=&\int_k Tr \left[\frac{1}{\slashed{k}+\slashed{k}_3-m}\slashed{p}\frac{1}{\slashed{k}+\slashed{k}_3+\slashed{p}-m}\gamma^{\mu}\frac{1}{\slashed{k}+\slashed{k}_3+\slashed{p}+\slashed{r}-m}\gamma^{\nu}\frac{1}{\slashed{k}+\slashed{k}_3+\slashed{p}+\slashed{Q}-m}\gamma^{\alpha}\gamma^{5}\right]\nonumber\\
&+\int_k Tr \left[\frac{1}{\slashed{k}+\slashed{k}_3-m}\gamma^{\mu}\frac{1}{\slashed{k}+\slashed{k}_3+\slashed{r}-m}\slashed{p}\frac{1}{\slashed{k}+\slashed{k}_3+\slashed{p}+\slashed{r}-m}\gamma^{\nu}\frac{1}{\slashed{k}+\slashed{k}_3+\slashed{p}+\slashed{Q}-m}\gamma^{\alpha}\gamma^{5}\right]\nonumber\\
&+\int_k Tr \left[\frac{1}{\slashed{k}+\slashed{k}_3-m}\gamma^{\mu}\frac{1}{\slashed{k}+\slashed{k}_3+\slashed{r}-m}\gamma^{\nu}\frac{1}{\slashed{k}+\slashed{k}_3+\slashed{Q}-m}\slashed{p}\frac{1}{\slashed{k}+\slashed{k}_3+\slashed{p}+\slashed{Q}-m}\gamma^{\alpha}\gamma^{5}\right]\label{bolha3}
\end{align}
where $k_i$ are arbitrary momentum routings and $Q=s+r$.

\begin{figure}[!h]
\includegraphics[trim=20mm 60mm 80mm 60mm,scale=0.4]{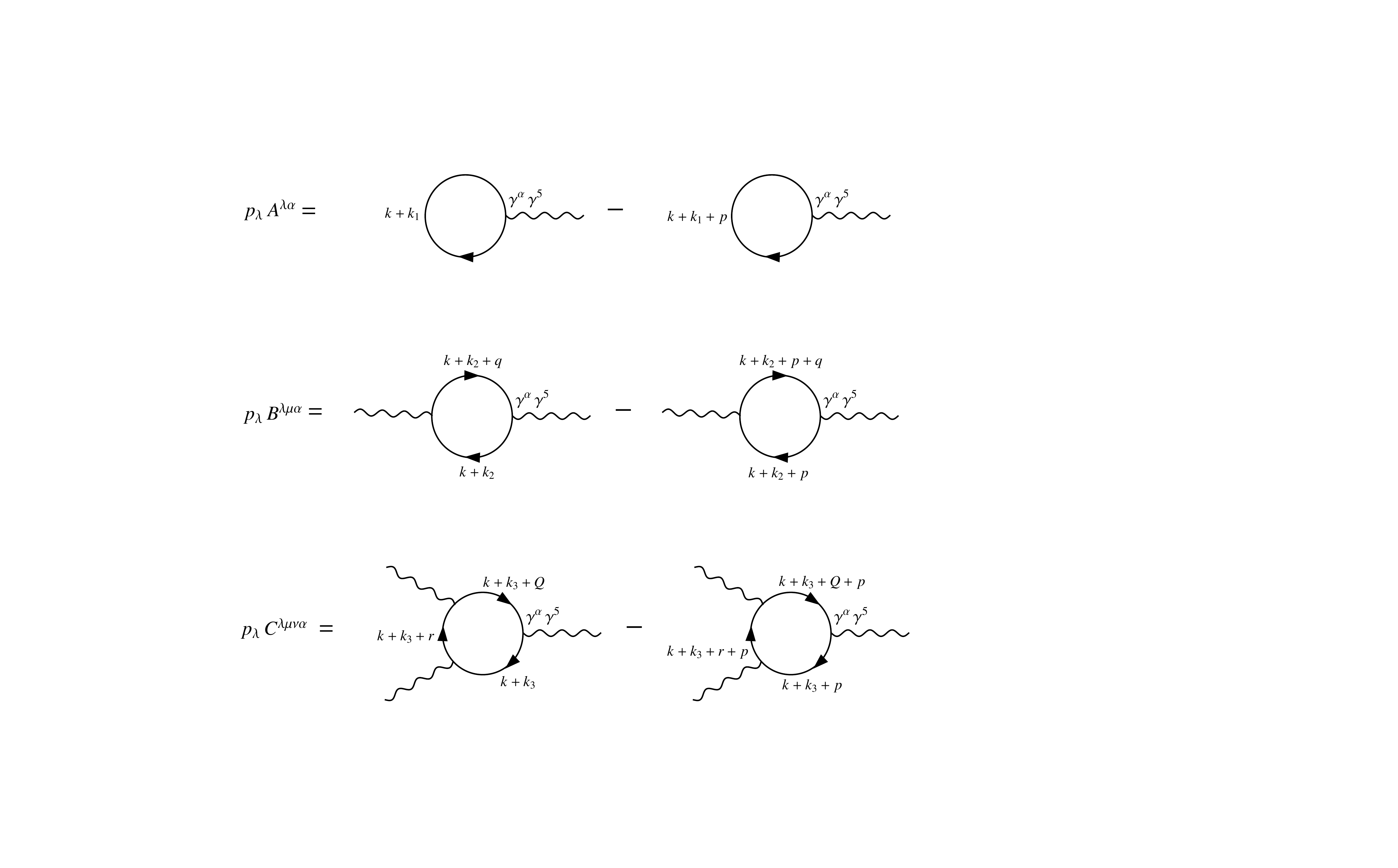}
\caption{Pictorial representation of the Ward identities, showing its connection to Momentum Routing Invariance.}
\label{FIGURA2}
\end{figure}

Diagrams with more than four external legs are finite and do not need to be considered. To proceed we apply the sum and subtraction of internal momenta, $\slashed{p}=(\slashed{k}+\slashed{k}_1+\slashed{p}-m)-(\slashed{k}+\slashed{k}_1-m)$, for example, which allows us to write
\begin{align}
&p_{\lambda}A^{\lambda\alpha}=\int_k Tr \left[\frac{1}{\slashed{k}+\slashed{k}_1-m}
\gamma^{\alpha}\gamma^{5}\right]-\int_k Tr \left[\frac{1}{\slashed{k}+\slashed{k}_1+\slashed{p}-m}\gamma^{\alpha}\gamma^{5}\right]
\\
p_{\lambda}B^{\lambda\mu\alpha}=&\int_k Tr \left[\frac{1}{\slashed{k}+\slashed{k}_2-m}\gamma^{\mu}\frac{1}{\slashed{k}+\slashed{k}_2+\slashed{q}-m}\gamma^{\alpha}\gamma^{5}\right]\nonumber\\
&-\int_k Tr \left[\frac{1}{\slashed{k}+\slashed{k}_2+\slashed{p}-m}\gamma^{\mu}\frac{1}{\slashed{k}+\slashed{k}_2+\slashed{q}+\slashed{p}-m}\gamma^{\alpha}\gamma^{5}\right]\\
p_{\lambda}C^{\lambda\mu\nu\alpha}=&\int_k Tr \left[\frac{1}{\slashed{k}+\slashed{k}_3-m}\gamma^{\mu}\frac{1}{\slashed{k}+\slashed{k}_3+\slashed{r}-m}\gamma^{\nu}\frac{1}{\slashed{k}+\slashed{k}_3+\slashed{Q}-m}\gamma^{\alpha}\gamma^{5}\right]\nonumber\\
&-\int_k Tr \left[\frac{1}{\slashed{k}+\slashed{k}_3+\slashed{p}-m}\gamma^{\mu}\frac{1}{\slashed{k}+\slashed{k}_3+\slashed{r}+\slashed{p}-m}\gamma^{\nu}\frac{1}{\slashed{k}+\slashed{k}_3+\slashed{p}+\slashed{Q}-m}\gamma^{\alpha}\gamma^{5}\right]
\end{align}
\noindent
whose pictorial representation can be found on fig. \ref{FIGURA2}.

Noticing that a Feynman diagram, in general, respects MRI if the difference of the same diagram with two different momentum routings vanishes, one can easily see that the RHS of the equation above is just the condition to implement MRI for the diagrams upon which the diagrammatic proof of gauge invariance is constructed. Therefore, it is clear from the pictorial representation above that MRI is intrinsically connected with vectorial gauge symmetry even in the case of effective Chiral Abelian Gauge Theories, meaning that the requirement of MRI is a sufficient condition to implement gauge symmetry and vice-versa.   

It should be emphasized that until now no regularization framework was assumed, showing that the connection between MRI and vectorial gauge symmetry holds in general. Therefore, any regularization that preserves (breaks) one symmetry will automatically preserve (break) the other. This connection is more clear in our framework since, as demonstrated in \cite{Cherchiglia}, the breaking of MRI (and consequently of gauge symmetry) is parametrized by surface terms which are regularization dependent and arbitrary by nature. This general feature can also be seen in our previous examples, in which the vectorial Ward identity was always proportional to surface terms only. For completeness we evaluate equations (\ref{bolha})-(\ref{bolha3}) in IReg which furnish
\begin{eqnarray}
p_{\lambda}A^{\lambda\alpha}&=&0, \label{WI1}\\
p_{\lambda}B^{\lambda\mu\alpha}&=&4i\upsilon_0 \epsilon^{\mu \alpha \lambda \beta}q_{\beta}p_{\lambda},\label{WI2}\\
p_{\lambda}C^{\lambda\mu\nu\alpha}&=&4i\upsilon_0 p_{\lambda}\epsilon^{\alpha\nu\mu\lambda},\label{WI3}
\label{WI5}
\end{eqnarray}
\noindent
showing that surface terms appear as expected. 

\begin{figure}[!h]
\includegraphics[trim=20mm 60mm 80mm 60mm,scale=0.4]{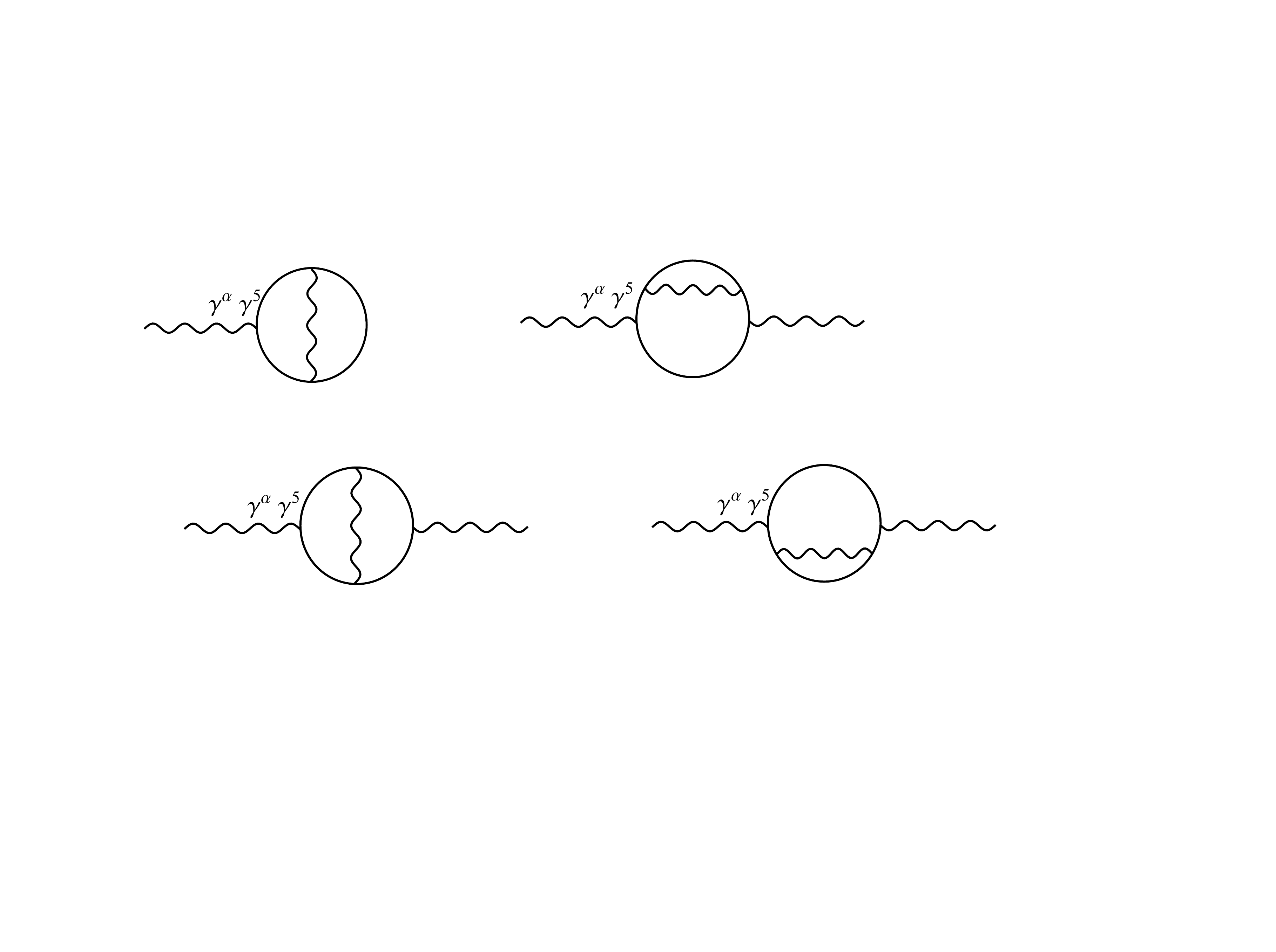}
\caption{One and two-point functions needed for the two-loop diagrammatic proof of gauge invariance.}
\label{FIGURA3}
\end{figure}

To conclude this section we discuss how the above proof can be extended to arbitrary loop order. We sketch first the diagrammatic proof for the two-loop case in some detail. As explained in \cite{Peskin}, the idea behind the diagrammatic proof of gauge invariance is the following: consider we have an arbitrary amplitude $\mathcal{M}_0$ with a closed fermionic loop. The Ward identity is then just obtained by inserting an external photon in all possible ways in the basic amplitude $\mathcal{M}_0$. This was our approach in the one-loop case in which the basic diagrams of fig. \ref{FIGURA1} gave rise to the pictorial representation of the Ward identities of fig. \ref{figd}. For the two-loop case the procedure is similar, and one needs first to depict all genuine two-loop corrections\footnote{By genuine we mean diagrams without one-loop closed fermion loops as sub-diagrams, since this case is the one we just studied.}. For simplicity, we just depict the one and two-point contributions in fig. \ref{FIGURA3}.

Our next task is to insert the external photon in all possible ways, giving rise to a pictorial representation of Ward identities as we did for the one-loop case. Explicitly, for the one-point function, we have

%
%
%
\begin{figure}[!h]
\includegraphics[trim=0mm 60mm 0mm 40mm,scale=0.5]{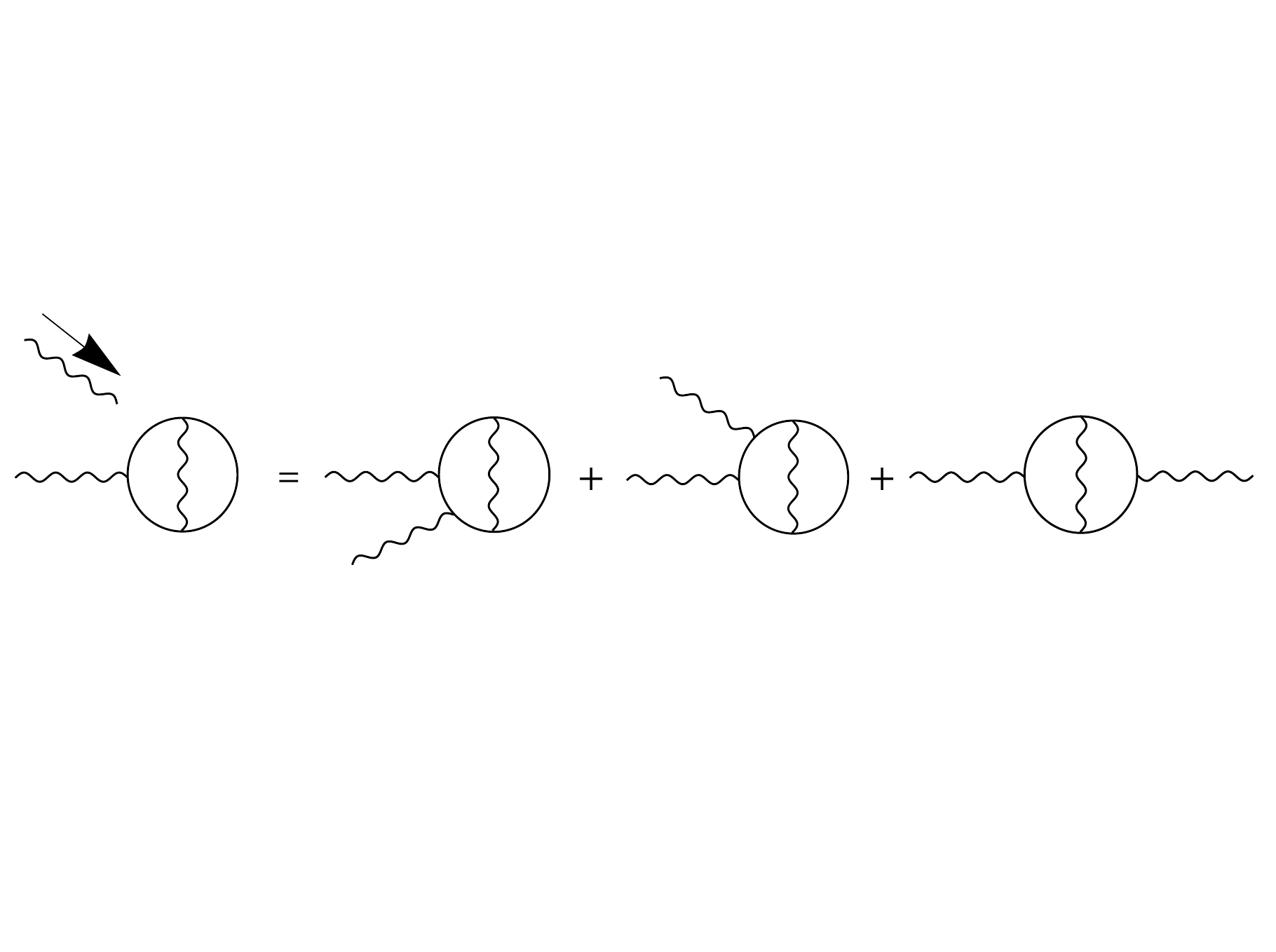}
\caption{Pictorial representation of two-loop Ward identity for the one-point function.}
\label{2loop}
\end{figure}

The corresponding amplitude of the first contribution of figure \ref{2loop} can be rewritten as below
\begin{align}
\int_{k_1}\int_{k_2} Tr \left[\frac{1}{\slashed{k}_1-m}\slashed{p}\frac{1}{\slashed{k}_1+\slashed{p}-m}\gamma^{\sigma}\frac{1}{\slashed
{k}_2+\slashed{p}-m}\gamma^{\rho}\frac{1}{\slashed{k}_1+\slashed{p}-m}\gamma^{\alpha}\gamma^{5}\right]\frac{g_{\sigma\rho
}}{(k_1-k_2)^2-m^2}= \nonumber\\ =\int_{k_1}\int_{k_2} Tr \left[\frac{1}{\slashed{k}_1-m}\gamma^{\sigma}\frac{1}{\slashed
{k}_2+\slashed{p}-m}\gamma^{\rho}\frac{1}{\slashed{k}_1+\slashed{p}-m}\gamma^{\alpha}\gamma^{5}\right]\frac{g_{\sigma\rho
}}{(k_1-k_2)^2-m^2}-\nonumber\\ -\int_{k_1}\int_{k_2} Tr \left[\frac{1}{\slashed{k}_1+\slashed{p}-m}\gamma^{\sigma}\frac{1}{\slashed
{k}_2+\slashed{p}-m}\gamma^{\rho}\frac{1}{\slashed{k}_1+\slashed{p}-m}\gamma^{\alpha}\gamma^{5}\right]\frac{g_{\sigma\rho
}}{(k_1-k_2)^2-m^2},
\end{align}
where we replace $\slashed{k}=(\slashed{k}+\slashed{p}_1-m)-(\slashed{p}_1-m)$ to get in the second line.

\begin{figure}[!h]
\includegraphics[trim=40mm 40mm 20mm 20mm,scale=0.4]{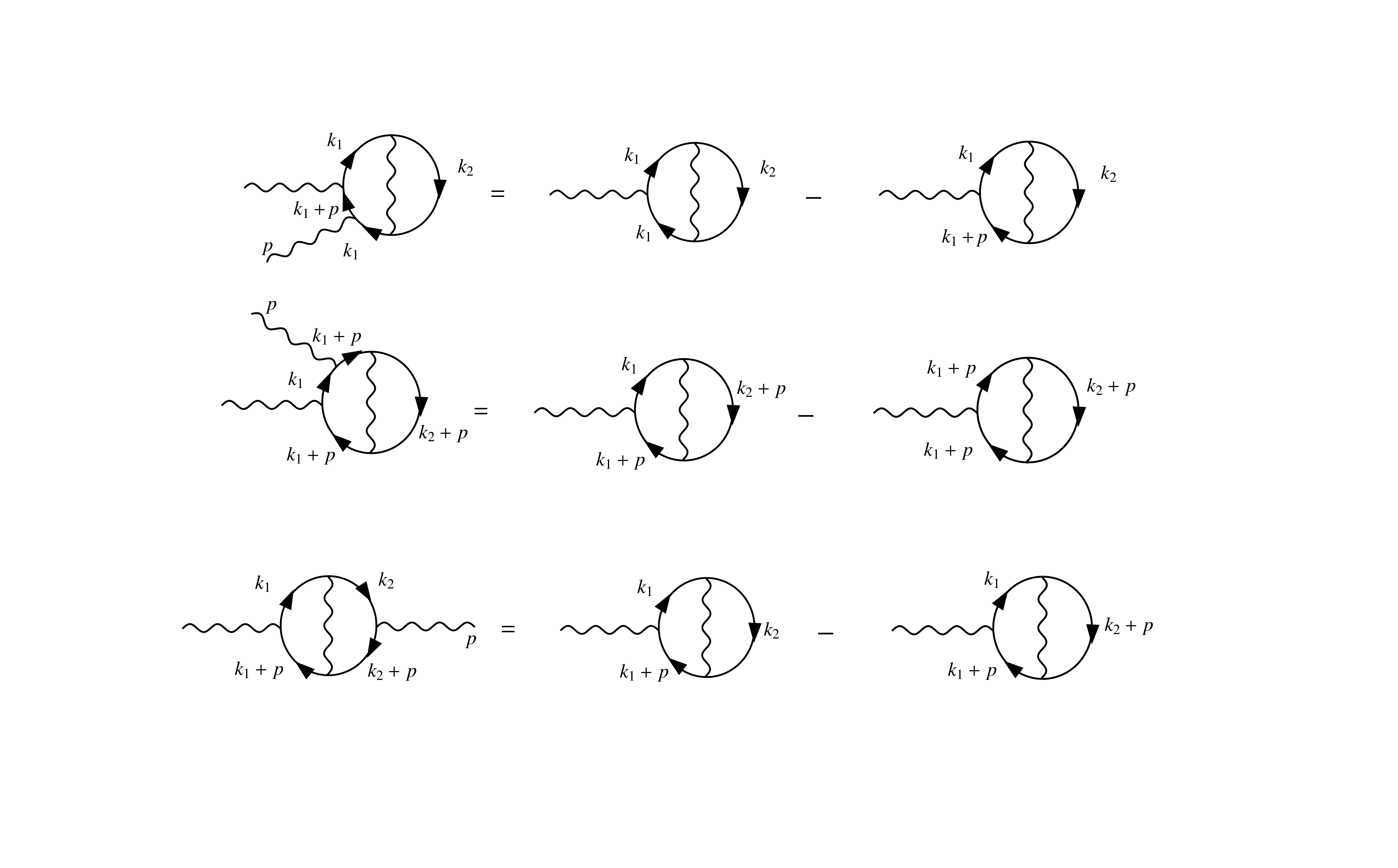}
\caption{Pictorial representation of the result of each tadpole insertion.}
\label{tad}
\end{figure}

In a similar fashion, we calculate the result of the other two insertions. We present the diagrammatic result of those insertions in 
figure \ref{tad}.

\vspace{-0.5cm}

\begin{figure}[!h]
\centering
    \subfigure[]
    {\includegraphics[trim=20mm 150mm 20mm 150mm,scale=0.4]{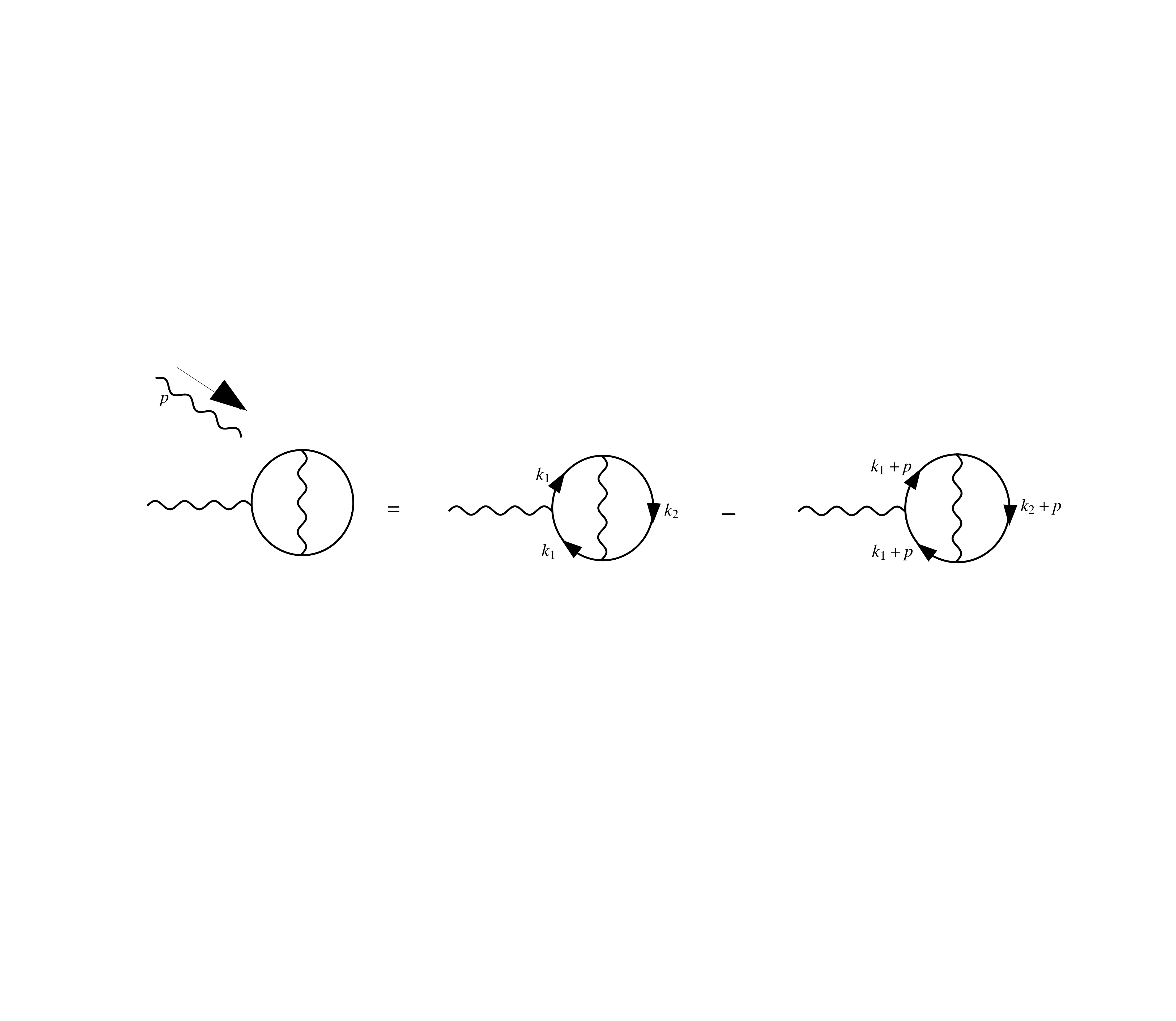}
        \label{diff}}											
		\subfigure[]
{  \includegraphics[trim=20mm 150mm 20mm 110mm,scale=0.4]{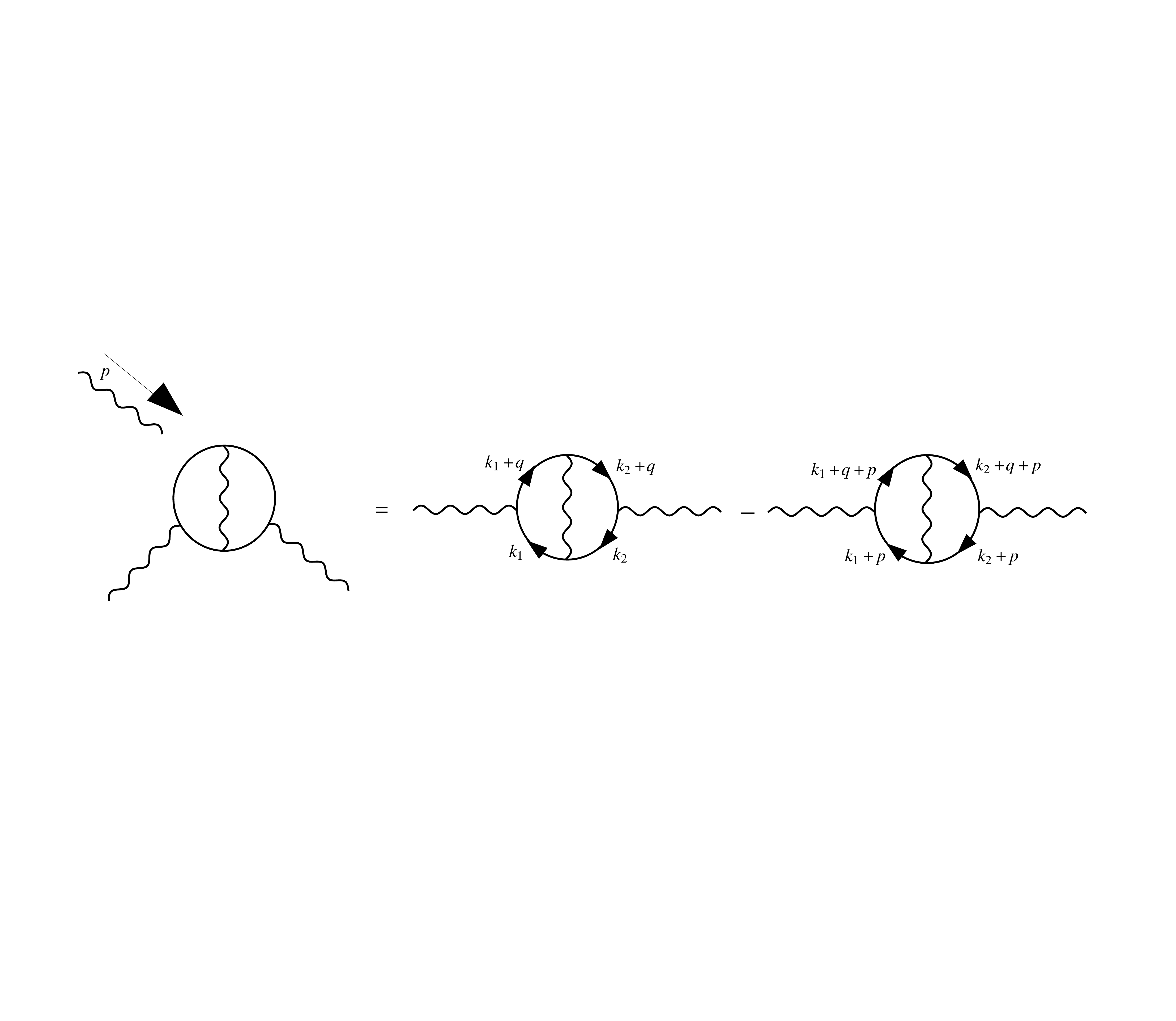}
        \label{diff2}}								
				\caption{ Pictorial representation of the relation between gauge and MRI for two- and three-point two loop diagrams, 
(a) and (b) respectively. The external momentum $p$ acts as an arbitrary routing and making the right-hand side zero is the MRI condition while making the left-hand side zero is the gauge invariance condition.}
\label{fig}
\end{figure}

Thus, by summing all contributions we obtain the result presented in figure \ref{diff}. 
This picture shows, for the two-loop case as well, that vectorial gauge symmetry is connected with MRI. Moreover, since MRI breaking is always parametrized by surface terms (as showed in \cite{Cherchiglia}), one can see that these ambiguities will also be connected with vectorial gauge breakings, in the sense that only by setting these terms to zero one obtains a momentum routing (and gauge) invariant result.

To complete the proof, one has to compute the other two Ward identities ($p_{\lambda}B^{\lambda\mu\alpha}$, $p_{\lambda}C^{\lambda\mu\nu\alpha}$) which have, respectively, the two-loop two- and three-point functions as building blocks. The calculation is straightforward, however, due to the lack of space, we just present the computation for one of the two-point functions depicted in fig. \ref{FIGURA3}. At first, one might expect that only the computation of the full sum is meaningful, however, as we are going to show,
%
%
%
%
%
 diagrams for $\mathcal{M}_0$ respect gauge invariance individually (if one sets surface terms equal to 
zero) and not only their sum. Adopting the same strategy for the one-point function, one obtains the result depicted in fig. \ref{diff2}, which shows, once again, that MRI is connected with the vectorial gauge symmetry. In a similar way, all other diagrams can be shown to present the same behavior found in figure \ref{fig}:
gauge invariance is implemented if, and only if, MRI is guaranteed. This proof can be extended in a straightforward way to an arbitrary
number of loops.


\begin{thebibliography}{99}

\bibitem{DR} G. 't Hooft and M. Veltman, {\it Nucl. Phys. B} {\bf 44} (1972) 189.
\bibitem{Bollini} C. G. Bollini and J. J. Giambiagi, {\it Nuovo Cimento} {\bf 12} (1972) 20.
\bibitem{Siegel} W. Siegel, {\it Phys. Lett. B} {\bf 84} (1979) 189.
\bibitem{Siegel2} W. Siegel, {\it Phys. Lett. B} {\bf 94} (1980) 37.
\bibitem{Jegerlehner} F. Jegerlehner, {\it Eur. Phys. J. C} {\bf 18} (2001) 673.
\bibitem{Tsai1} E. C. Tsai, {\it Phys. Rev. D} {\bf 83} (2011) 025020.
\bibitem{Tsai2} E. C. Tsai, {\it Phys. Rev. D} {\bf 83} (2011) 065011.
\bibitem{Ferrari1} R. Ferrari, 2014, hep-th/1403.4212.
\bibitem{Ferrari2} R. Ferrari, 2015, hep-th/1503.07410.
\bibitem{Ferrari3} R. Ferrari, 2016, hep-th/1605.06929.
\bibitem{Cynolter} G. Cynolter and E. Lendvai, {\it Mod. Phys.  Lett. A} {\bf 26} (2011) 1537.
\bibitem{IR} O. A. Battistel, A. L. Mota and M. C. Nemes, {\it Mod. Phys. Lett. A} {\bf 13} (1998)1597.
\bibitem{BaetaScarpelli:2000zs}  A. P. Baeta Scarpelli, M. Sampaio and M. C. Nemes,  {\it Phys. Rev.  D} {\bf 63} (2001) 046004.
\bibitem{Cherchiglia:2010yd} A. L. Cherchiglia, M. Sampaio and M. C. Nemes, {\it Int. J. Mod. Phys. A} {\bf 26} (2011) 2591.
\bibitem{Cherchiglia} L. C. Ferreira, A. L. Cherchiglia, Brigitte Hiller, Marcos Sampaio and M. C. Nemes, {\it Phys. Rev. D} {\bf 86} (2012) 025016.
\bibitem{Cherchiglia:2012zp} A. L. Cherchiglia, L. A. Cabral, M. C. Nemes and M. Sampaio, {\it Phys. Rev. D} {\bf 87} (2013) no.6,  065011.
\bibitem{Felipe:2014gma}  J. C. C. Felipe, A. R. Vieira, A. L. Cherchiglia, A. P. Baêta Scarpelli and M. Sampaio, {\it Phys. Rev. D} {\bf 89} (2014) no.10,  105034.
\bibitem{Vieira:2015fra} A. R. Vieira, A. L. Cherchiglia and M. Sampaio, {\it Phys. Rev. D} {\bf 93} (2016) no.2,  025029.
\bibitem{Jackiw} R. Jackiw and R. Rajaraman, {\it Phys. Rev. Lett.} {\bf 54} (1985) 1219.
\bibitem{JackiwFU} R. Jackiw  , \textit{Int. J. Mod. Phys. B} \textbf{14} (2000) 2011.
\bibitem{Jacquot:1999hj} J.~L.~Jacquot, Eur.\ Phys.\ J.\ C {\bf 23} (2002) 349.
\bibitem{Bonneau} G. Bonneau, {\it Phys. Lett. B} {\bf 96} (1980) 147.
\bibitem{Thompson} G. Thompson and H. -L. Yu, {\it Phys. Lett. B} {\bf 151} (1985) 119.
\bibitem{Baikov} P. A. Baikov and V. A. Ilyin, {\it Theor. Math. Phys.} {\bf 88} (1991) 789.
\bibitem{Zralek} M. I. Abdelhafiz and M. Zralek, {\it Acta Phys. Pol. B} {\bf 18} (1987) 21.
\bibitem{BellJackiw} J. S. Bell and R. Jackiw, {\it Nuovo Cimento A} {\bf 60} (1969) 47.
\bibitem{Adler} S. L. Adler, {\it Phys. Rev.} {\bf 177} (1969) 2426.
\bibitem{Elias} V. Elias, G. McKeon and R. B. Mann, {\it Nucl. Phys. B} {\bf 229} (1983) 487.
\bibitem{YU} H.-L. Yu and W. B. Yeung, {\it Phys. Rev. D} {\bf 35} (1987) 3955.
\bibitem{Bertlmann} R. A. Bertlmann, {\it Anomalies in Quantum Field Theory}, International Series of Monographs in Physics 
(Oxford University Press, Oxford, 1996).
\bibitem{Fujikawa} K. Fujikawa, {\it Phys. Rev. Lett.} {\bf 42} (1979) 1195.
\bibitem{Zee} B. Zumino, W. Yong-Shi and A. Zee, {\it Nucl. Phys. B} {\bf 239} (1984) 477.
\bibitem{tHooft} G. t'Hooft, {\it Phys. Rev. Lett.} {\bf 37} (1976) 8.
\bibitem{Perez} F. del Aguila and M. Perez-Victoria, {\it Acta Phys. Pol. B} {\bf 29} (1998) 2857.
\bibitem{Wu} Y. L. Ma and Y. L. Wu, {\it Int. J. Mod. Phys. A} {\bf 21} (2006) 6383.
\bibitem{Scarpelli} A. P. B. Scarpelli, M. Sampaio, M. C. Nemes and B. Hiller, {\it Phys. Rev. D} {\bf 
64} (2001) 046013.
\bibitem{Peskin} M. E. Peskin and D. Schroeder, {\it An Introduction to Quantum Field Theory} (Addison-Wesley, Reading MA, 1995).
\bibitem{Orimar} O. A. Battistel and G. Dallabona, {\it Eur. Phys. J. C} {\bf 45} (2006) 721.
\bibitem{Carneiro:2003id}
  D.~E.~Carneiro, A.~P.~Baeta Scarpelli, M.~Sampaio and M.~C.~Nemes,
  JHEP {\bf 0312} (2003) 044

\bibitem{Fargnoli:2010mf}
  H.~G.~Fargnoli, B.~Hiller, A.~P.~B.~Scarpelli, M.~Sampaio and M.~C.~Nemes,
  Eur.\ Phys.\ J.\ C {\bf 71} (2011) 1633

\bibitem{Cherchiglia:2015vaa}
  A.~L.~Cherchiglia, M.~Sampaio, B.~Hiller and A.~P.~B.~Scarpelli,
  Eur.\ Phys.\ J.\ C {\bf 76} (2016) no.2,  47

\bibitem{SUSY}
 A.~C.~D.~Viglioni, A.~L.~Cherchiglia, A.~R.~Vieira, B.~Hiller and M.~Sampaio,
work in progress.
\end{thebibliography}

\section{Concluding remarks}
\label{conclusion}

In this contribution we intended to shed some light on chiral anomalies, their connection to momentum routing invariance and how a democratic framework for 
Ward identities (vectorial and axial, respectively) can be constructed. Particularly, we studied two and four dimensional theories and have shown that, relying 
on the symmetrization of traces containing dimension specific objects such as $\gamma_{5}$, the Ward identities can be displayed in a democratic way. In our 
view this is the most natural approach to be followed since the Ward identity to be (or not) preserved should result from the Physics and/or phenomenology requirements, and not be an outcome conditioned by the regularization method applied.

In this context, we have also studied momentum routing dependence for effective abelian chiral theories. We have shown that, as in the case of non-chiral theories, momentum routing 
independence (MRI) is achieved if we set surface terms (which represent ambiguities and allow a democratic display for Ward identities) to zero. We also have shown that, even for effective chiral theories, the vectorial gauge invariance is guaranteed if, and only if, we set surface terms to zero. Since this is the same 
condition to implement MRI, we can conclude that both symmetries are intrinsically connected. 

As perspectives, we remark the application of the minimal prescription here presented to deal with dimension specific objects such as $\gamma_{5}$ to other contexts as well, for instance, supersymmetric theories. Some investigations in this direction, connecting regularization ambiguities with supersymmetry breaking, have already been done by some of us in \cite{Carneiro:2003id,Fargnoli:2010mf,Cherchiglia,Cherchiglia:2015vaa}. A more complete investigation is under progress \cite{SUSY}. We also remark that the prescription here presented, connecting regularization ambiguities with the breaking of vectorial gauge invariance in general, allows a systematic application of IReg to effective chiral theories in general. Therefore, one avoids tedious checks and addition of symmetry restoring counterterms as requested by other methods. 

\section*{Acknowledgments}

A. Cherchiglia acknowledges fruitful discussions with D. St\"{o}ckinger and M. P\'{e}rez-Victoria. 
A. Cherchiglia, A. Vieira and M. S. acknowledge financial support by CNPq, Conselho Nacional de Desenvolvimento Cient\'{\i}fico e Tecnol\'{o}gico - Brazil. M. S. acknowledges financial support by FAPEMIG. 

\section*{Appendix} \label{A}

We perform the computation of the finite part of the triangle diagram, $T^{finite}_{\mu\nu\alpha}$. Since it does not
depend on the routing, we can choose $k_1=0$, $k_2=q$ e $k_3=-p$ and we have:

\begin{align}
T_{\mu\nu\alpha}=&-i\int_k Tr\left[\gamma_{\mu}\frac{i}{\slashed{k}-m}\gamma_{\nu}\frac{i}{\slashed{k}+\slashed
{q}-m}\gamma_{\alpha}\gamma_5 \frac{i}{\slashed{k}-\slashed{p}-m}\right]+(\mu \leftrightarrow \nu, p\leftrightarrow q)=\nonumber\\
&= -8i\upsilon_0 \epsilon_{\mu \nu \alpha \beta}(q-p)^{\beta}+T^{finite}_{\mu\nu\alpha}
\end{align}

After taking the trace and regularizing we find out the finite part of the amplitude. We list the results of the integrals in the final
part of this section. The result is

\begin{align}
T^{finite}_{\mu\nu\alpha}=& 4i b\{\epsilon_{\alpha\mu\nu\lambda}q^{\lambda}(p^2\xi_{01}(p,q)-q^2\xi_{10}(p,q))+\epsilon_{
\alpha\mu\nu\lambda}q^{\lambda}(1+2m^2\xi_{00}(p,q))+
\nonumber\\ & +4\epsilon_{\alpha\nu\lambda\tau}p^{\lambda}q^{\tau}[(\xi_{01}(p,q)-\xi_{02}(p,q))p_{\mu}+\xi_{11}(p,q)q_{\mu}]+
(\mu \leftrightarrow \nu, p\leftrightarrow q)\},
\label{Tfinito}
\end{align}
where the functions $\xi_{nm}(p,q)$ are defined as
\be
\xi_{nm}(p,q)=\int^1_0 dz\int^{1-z}_0 dy \frac{z^n y^m}{Q(y,z)},\\
\ee
with
\be
Q(y,z)=[p^2 y(1-y)+q^2 z(1-z)+2(p\cdot q)yz-m^2]
\ee
and those functions have the property $\xi_{nm}(p,q)=\xi_{mn}(q,p)$.

Those integrals obey the following relations which we have already used in the derivation of equation (\ref{Tfinito})
\bq
&q^2 \xi_{11}(p,q)-(p\cdot q)\xi_{02}(p,q)=\frac{1}{2}\left[ -\frac{1}{2}Z_0((p+q)^2,m^2)+\frac{1}{2}Z_0(p^2,m^2)+q^2 \xi_{01}(p,q)
\right],\label{f1}\\
&p^2 \xi_{11}(p,q)-(p\cdot q)\xi_{20}(p,q)=\frac{1}{2}\left[ -\frac{1}{2}Z_0((p+q)^2,m^2)+\frac{1}{2}Z_0(q^2,m^2)+p^2 \xi_{10}(p,q)
\right],\label{f2}\\
&q^2 \xi_{10}(p,q)-(p\cdot q)\xi_{01}(p,q)=\frac{1}{2}[ -Z_0((p+q)^2,m^2)+Z_0(p^2,m^2)+q^2 \xi_{00}(p,q)],\label{f3}\\
&p^2 \xi_{01}(p,q)-(p\cdot q)\xi_{10}(p,q)=\frac{1}{2}[ -Z_0((p+q)^2,m^2)+Z_0(q^2,m^2)+p^2 \xi_{00}(p,q)],\label{f4}\\
&q^2 \xi_{20}(p,q)-(p\cdot q)\xi_{11}(p,q)=\frac{1}{2}\left[-\left(\frac{1}{2}+m^2\xi_{00}(p,q)\right)+\frac{1}{2}p^2\xi_{01}(p,q)+\frac
{3}{2}q^2\xi_{10}(p,q)\right],\label{f5}\\
&p^2 \xi_{02}(p,q)-(p\cdot q)\xi_{11}(p,q)=\frac{1}{2}\left[-\left(\frac{1}{2}+m^2\xi_{00}(p,q)\right)+\frac{1}{2}q^2\xi_{10}(p,q)+\frac
{3}{2}p^2\xi_{01}(p,q)\right],\label{f6}
\eq
where $Z_k(p^2,m^2)$ is defined as
\be
Z_k(p^2,m^2)=\int^1_0 dz z^k \ln \frac{m^2-p^2 z(1-z)}{m^2}.
\ee

The derivation of the relations(\ref{f1})-(\ref{f6}) can be simply achieved by integration by parts. There is a whole review \cite
{Orimar} about those four dimensional integrals where we can find these details.

\subsection*{Result of the integrals of section \ref{Schwinger}}
\be
\int_q \frac{1}{(q^2-\mu^2)((q-p)^2-\mu^2)}=-\frac{2}{p^2}b \ln \left(\frac{-p^2}{\mu^2}\right),
\ee
\be
\int_q \frac{q^{\alpha}}{(q^2-\mu^2)((q-p)^2-\mu^2)}=-\frac{p^{\alpha}}{p^2}b \ln \left(\frac{-p^2}{\mu^2}\right),
\ee
\bq
\int_q \frac{q^{\alpha}q^{\beta}}{(q^2-\mu^2)((q-p)^2-\mu^2)}=&\frac{1}{2}g^{\alpha\beta}(I_{log}(\mu^2)-\upsilon_0 )-
\frac{b}{p^2}(g^{\alpha\beta}p^2-p^{\alpha}p^{\beta})\left(1-\frac{1}{2}\ln \left(\frac{-p^2}{\mu^2}\right)\right)-\nonumber\\&
-\frac{p^{\alpha}p^{\beta}}{2p^2}b \ln \left(\frac{-p^2}{\mu^2}\right),
\eq
where $\int_q\equiv \int^{\Lambda} \frac{d^2 q}{(2\pi)^2}$, $b=\frac{i}{4\pi}$ and $\mu$ is an infrared regulator.

\subsection*{Result of the integrals of section \ref{ABJ}}

\begin{equation}
\int_k \frac{1}{(k^2-m^2)[(k-p)^2-m^2][(k+q)^2-m^2]}= b \xi_{00}(p,q),
\end{equation}
\begin{equation}
\int_k \frac{k^{\alpha}}{(k^2-m^2)[(k-p)^2-m^2][(k+q)^2-m^2]}= b (p^{\alpha}\xi_{01}(p,q)-q^{\alpha}\xi_{10}(p,q)),
\end{equation}
\begin{align}
\centering
\int_k \frac{k^{2}}{(k^2-m^2)[(k-p)^2-m^2][(k+q)^2-m^2]}=& I_{log}(m^2)-b Z_0(q^2,m^2)+ b (m^2-p^2)\xi_{00}(p,q)+\nonumber\\
&+2b(p^2\xi_{01}(p,q)-(p\cdot q) \xi_{10}(p,q)),
\end{align}
\begin{align}
&\int_k \frac{k^{\alpha}k^{\beta}}{(k^2-m^2)[(k-p)^2-m^2][(k+q)^2-m^2]}= \frac{1}{4}g^{\alpha\beta}(I_{log}(m^2)-\upsilon_0)-\frac
{1}{4}b g^{\alpha\beta} Z_0(q^2,m^2)-\nonumber\\&-b\Big[\frac{1}{2}g^{\alpha\beta}p^2(\xi_{00}(p,q)-3\xi_{01}(p,q)-\xi_{10}(p,q)+2\xi_
{02}(p,q)+2\xi_{11}(p,q))-\xi_{02}(p,q)p^{\alpha}p^{\beta}+\nonumber\\&+\xi_{11}(p,q)q^{\alpha}p^{\beta}+\xi_{11}(p,q)p^{\alpha}q^{\beta}
-\xi_{20}(p,q)q^{\alpha}q^{\beta}+(\xi_{10}(p,q)-\xi_{11}(p,q)-\xi_{20}(p,q))g^{\alpha\beta}(p\cdot q)\Big],
\end{align}
\begin{align}
&\int_k \frac{k^{\alpha}k^{2}}{(k^2-m^2)[(k-p)^2-m^2][(k+q)^2-m^2]}=\frac{1}{2}(p^{\alpha}-q^{\alpha})(I_{log}(m^2)-\upsilon_0)+
\frac{1}{2}b(q^{\alpha}Z_0(q^2,m^2)-\nonumber\\ & -p^{\alpha}Z_0(p^2,m^2))+ b(m^2- q^2)(p^{\alpha}\xi_{01}(p,q)-q^{\alpha}\xi_{10}(p,q))+
b[q^{\alpha}p^2(\xi_{00}(p,q)-3\xi_{01}(p,q)-\xi_{10}(p,q)+\nonumber\\ &+2\xi_{02}(p,q)+2\xi_{11}(p,q))-2(p\cdot q)p^{\alpha}\xi_{02}(p,q)
+2 q^2 p^{\alpha}\xi_{11}(p,q)+2(p\cdot q)q^{\alpha}(\xi_{10}(p,q)-\nonumber\\&-\xi_{20}(p,q))-2 q^2 q^{\alpha}\xi_{20}(p,q))],
\end{align}
where  $\int_k\equiv \int^{\Lambda} \frac{d^4 k}{(2\pi)^4}$ and $b=\frac{i}{(4\pi)^2}$.

\end{document}